\documentclass[aps,showpacs,graphicx]{revtex4} 

\usepackage{amssymb} 
\usepackage[dvips]{graphicx} 
 \usepackage{hyperref}
\usepackage[english]{babel} 
\usepackage{braket}
\newcommand{\beq}{\begin{equation}} 
\newcommand{\eeq}{\end{equation}} 
\newcommand{\beqn}{\begin{eqnarray}} 
\newcommand{\eeqn}{\end{eqnarray}} 
\newcommand{\ds}{\displaystyle} 
\newcommand{\bsigma}{\mbox{\boldmath $\sigma$}} 
\newcommand{\btau}{\mbox{\boldmath $\tau$}} 
\newcommand{\half}{\frac{1}{2}} 

\newcommand{\br}{{\bf r}}

\newcommand{\threej}[6]{ \left( \begin{array}{ccc} 
                               #1 & #2 & #3 \\ 
                               #4 & #5 & #6  
                             \end{array} 
                        \right) }  
\newcommand{\sixj}[6]{ \left\{ \begin{array}{ccc} 
                               #1 & #2 & #3 \\ 
                                #4 & #5 & #6  
                               \end{array} 
                        \right\} }  
\newcommand{\ninej}[9]{ \left\{ \begin{array}{ccc} 
                               #1 & #2 & #3 \\ 
                                #4 & #5 & #6 \\  
                                #7 & #8 & #9  
                                \end{array} 
                        \right\} }  
\usepackage[usenames]{color} 
\usepackage{ulem}

\begin{document} 
 
\noindent 
\title{Tensor force and deformation in even-even nuclei} 
 
\author{G. Co'$^{\,1,2}$, M. Anguiano$^{\,3}$, A. M. Lallena$^{\,3}$ }
\affiliation{$^1$ Dipartimento di Matematica e Fisica ``E. De Giorgi'', 
  Universit\`a del Salento, I-73100 Lecce, ITALY, \\ 
$^2$ INFN Sezione di Lecce, Via Arnesano, I-73100 Lecce, ITALY, \\ 
$^3$ Departamento de F\'\i sica At\'omica, Molecular y 
  Nuclear, Universidad de Granada, E-18071 Granada, SPAIN
}  

\date{\today}

\bigskip 
 
\begin{abstract} 
The variational principle is used to build a model which describes open shell nuclei with ground state deformations. Hartree-Fock equations are solved by using single particle wave functions whose radial parts depend on 
the projection of the angular momentum on the quantization axis. Pairing effects are taken into account by solving Bardeen-Cooper-Schrieffer equations in each step of the minimisation procedure. The Gogny D1S finite-range interaction and an extension of it that includes tensor terms  are consistently used in both parts of our calculations. The model is applied to 
study a set of isotopes with 34 protons and of isotones with 34 neutrons. 
Total energies, density distributions, their radii and single particle energies are analysed and the results of our calculations are compared with 
the available
experimental data. We focused our attention on the effects of the deformation 
and of the tensor force on these observables. Our model describes open shell 
nuclei from a peculiar perspective and opens the possibility of 
future theoretical developments.  
\end{abstract} 
 
\maketitle 

\section{Introduction} 
\label{sec:intro} 

The basic approach to describe nuclear deformations in terms  of nucleons 
is the Nilsson model \cite{nil55}. In this model, the nucleons move independently of each other in a deformed potential and the single particle (s.p.) wave functions depend on the projection of the corresponding angular momenta on the quantization $z$-axis. For a fixed value of the s.p. angular momentum, the states with smaller absolute values of this projection are more bound  in case of prolate deformations, and the inverse happens in oblate nuclei \cite{boh75,rin80}.

While in the Nilsson model the deformed potential is an external input, in the model we present here this potential is obtained by considering an effective nucleon-nucleon interaction and the variational principle. By searching for the minimum of the energy functional in the Hilbert sub-space formed by Slater determinants, a set of Hartree-Fock (HF) equations is obtained \cite{rin80} and we make the s.p. wave functions used to build up these Slater determinant explicitly dependent on the $z$-projection of their total angular momentum. In the minimisation procedure, we  take care of the pairing interaction by carrying out Bardeen-Cooper-Schrieffer (BCS) calculations that modify the occupation probabilities of the s.p. states. In our model, which we call HFBCS, the deformation emerges because not all the s.p. states with the same total angular momentum, but with different $z$-axis projection, are occupied. 

In this article, we present our model and apply it to the description of the ground state of even-even nuclei. As a testing ground, we have considered a set of medium-heavy nuclei where the occupation of the s.p. states 
ends in the $f$-$p$ shell. Specifically, we have studied a set of Se isotopes, they have 34 protons, and a set of isotones with 34 neutrons. We focused our attention to the emergence of the deformations and, since we have considered effective nucleon-nucleon interactions which include tensor terms, on the role of these tensor terms .
 
The results obtained within the HFBCS model for the aforementioned even-even nuclei have been compared with those found in deformed Hartree-Fock-Bogoliubov (HFB) calculations and with the available empirical values of their binding energies and their charge radii and distributions. In addition, also the angular momenta of the neighboring odd-even nuclei have been 
analyzed. The relevance of the use of this new set of s.p. wave functions 
is studied by comparing our results with those of the spherical HF+BCS model of Ref. \cite{ang14}. 

We present in Sec.~\ref{sec:model} the theoretical background of our model, and, in Sec.~\ref{sec:calculations}, the technical details of our calculations. In Sec. \ref{sec:results} we show, and discuss, the results we have obtained, focusing on deformations and the role of the tensor force. 
The conclusions of our study are summarized in Sec. \ref{sec:conclusions}.

\section{The model} 
\label{sec:model} 

In our description of an even-even nucleus, composed by $A$ nucleons, the 
basic ingredient is the set of s.p. wave functions used to build the Slater determinants. We assume that these s.p. wave functions, $\phi_k(x)$, can be factorized as:
\beq
\phi_k(x) \,= \, {R}_{k}{(r)} \ket{\tilde{k}} \chi_{t_k} \, ,
\label{eq:spwave} 
\eeq
where we have indicated with $x$ the generalized coordinate, which includes the position $\br$ with respect to the nuclear centre, the spin and the isospin of the considered nucleon.

The radial part of the s.p. wave function,
\beq
{R}_{k}{(r)}\,\equiv \, R_{n_k l_k j_k,\,m_k}^{t_k}(r) \, ,
\label{eq:spwave-rad} 
\eeq
is a function of $r\equiv | \br |$ and depends not only on the principal quantum number $n_k$, on the orbital orbital angular momentum quantum number $l_k$, on the total angular momentum quantum number $j_k$, and on the 
isospin third component $t_k$, but also on the  projection of $j_k$ on the $z$-axis, $m_k$. 
The part of the s.p. wave function depending on the angular coordinates, $\Omega_k \equiv(\theta_k,\phi_k)$, and on the spin third component, $s_k$, is:
\beq
\ket{\tilde{k}} \, \equiv \, \ket{l_k \half j_k m_k } \,=\, 
\sum_{\mu_k s_k} \braket{l_k \mu_k \half s_k | j_k m_k}
Y_{l_k \mu_k}(\Omega_k) \,\chi_{s_k} \, ,
\label{eq:spwave-ang} 
\eeq
where $Y_{l_k \mu_k}$ is a spherical harmonic, the symbol $\braket{|}$ indicates a Clebsch-Gordan coefficient, and $\chi_{s_k}$ is a Pauli spinor. 
Finally, in Eq.~(\ref{eq:spwave}), $\chi_{t_k}$ indicates the Pauli spinor related to the isospin. 

We assume time-reversal invariance \cite{boh75}; this means that, in our approach, ${R}^{t_k}_{n_k l_k j_k,\, m_k}{(r)} =  {R}^{t_k}_{n_k l_k j_k,-m_k}{(r)}$, and each  s.p. level is two-fold degenerated. 
This also implies that the nucleus has the shape of an ellipsoid whose symmetry axis is the $z$-axis.

We build a Slater determinant $\Phi$ with the s.p. wave functions $\phi_k$ of Eq.~(\ref{eq:spwave}), and the application of the variational principle to search for the minimum of the corresponding energy functional, $E[\Phi]$, leads to a set of differential equations of the type:
\beq
\nonumber 
\left[ \braket{\tilde{k} |-\frac{\hbar^2}{2 m} \nabla^2 |\tilde{k}} + \, 
{\cal U}_k(r_1) \,+ \, {\cal K}(r_1) \right] R_{k}(r_1) 
\,-\,  \int {\rm d}r_2 \, r_2^2 \, {\cal W}_k(r_1,r_2) \, R_k(r_2) \,  = 
\, \epsilon _{k} \, R_{k}(r_1) \, .
\label{eq:hfradial}
\eeq
As indicated by the first term, that of the kinetic energy, this expression has been obtained by integrating, and summing, on the angular, and spin, coordinates. 
In the above equation, $\epsilon_k$ indicates the s.p. energy, 
and the terms  depending on the two-body effective nucleon-nucleon interaction, 
$V(\br_1,\br_2)$, are the so-called Hartree potential,
\beq
{\cal U}_k(r_1) \, =\,  \sum_{i=1}^A \, v^2_i \int {\rm d}r_2 \, r_2^2 \,
R^2_{i}(r_2) \braket{\tilde{k} \tilde{i} |V(\br_1,\br_2)| \tilde{k} \tilde{i} } \, ,
\label{eq:U}
\eeq
the  Fock-Dirac term, 
\beq
{\cal W}_k(r_1,r_2)\, = \, \sum_{i=1}^A\, v^2_i \, \left[ R_{i}^*(r_2) \, R_{i}(r_1) 
\braket{\tilde{k} \tilde{i} |V(\br_1,\br_2)| \tilde{i} \tilde{k} }  
\right] \, ,
\label{eq:W}
\eeq
and a term related to the density dependence of the interaction:
\beqn
\nonumber
{\cal K}(r_1) \,=\,\frac{1}{4 \pi} \,\sum_{i,j=1}^A \, v^2_i \, v^2_j 
\int {\rm d}r_2 \, r_2^2 \, 
\left[ R_{i}^*(r_1) \, R_{j}^*(r_2)\braket{\tilde{i} \tilde{j} |\frac{\partial V(\br_1,\br_2)} {\partial \rho} | \tilde{i} \tilde{j} } R_{i}(r_1)  
\, R_{j}(r_2) \right. &&\\ && \left.
\hspace*{-8.5cm}
-\, R_{i}^*(r_1) \, R_{j}^*(r_2)  \braket{\tilde{i} \tilde{j} |\frac{\partial V(\br_1,\br_2)}  {\partial \rho} | \tilde{j} \tilde{i} } R_{j}(r_1) \, R_{i}(r_2)  \right] \, .
\label{eq:K}
\eeqn

The effective force $V(\br_1,\br_2)$ used in our calculations is a finite-range interaction that includes the four traditional central terms  (scalar, isospin, spin and spin-isospin), a zero-range spin-orbit term, a scalar density dependent term, and a tensor and a tensor-isospin dependent terms. More explicit expressions of ${\cal U}$, ${\cal W}$ and ${\cal K}$ are presented in Appendix~\ref{sec:hfmatel}.

In Eqs. (\ref{eq:U})-(\ref{eq:K}), we have indicated with $v_k^2$ the occupation probabilities of the s.p. states. Their values are obtained by solving the set of BCS equations:
\beq
v^2_{k} = \half \left[ 1 \,- \, \frac{\epsilon_{k} \,- \, \lambda} {(\epsilon_{k} \,- \, \lambda)^2 \, - \, \Delta^2_{k} } \right] \, ,
\label{eq:v2}
\eeq
where $\lambda$, the energy gap, is given by:
\begin{equation} 
\lambda \,=\, \frac {\displaystyle 2\, 
\sum_{k} v^2_{k}  \,-\, \sum_{k} \left( 1 \,-\, \frac{\epsilon_{k}}{ \sqrt{ \epsilon^2_{k} + \Delta^2_{k}}} \right)  }
{\displaystyle \sum_{k} \frac{\epsilon_{k}}{ \sqrt{ \epsilon^2_{k} + \Delta^2_{k}}}} \, .
\label{eq:lambda}
\end{equation}
The quantity $\Delta_k$ satisfies the relation
\beq
\Delta_{k} = - \half \sum_{i} 
\frac{\Delta_{i}}{\sqrt{ (\epsilon_{i} - \lambda)^2 +\Delta^2_{i}}} \braket{k k\, 00 |V(\br_1,\br_2) | i i \, 00}  \, ,
\label{eq:delta} 
\eeq
with $\ket{\alpha \alpha \,00}$ indicating a state where the s.p. states $\phi_\alpha$ are coupled to total angular momentum $J=0$ and $z$-axis projection $M=0$. 

The set of Eqs. (\ref{eq:v2})-(\ref{eq:delta}) are the BCS equations. The 
effective nucleon-nucleon interaction enters in the matrix element of Eq.~(\ref{eq:delta}) whose detailed expression is presented in Appendix \ref{sec:bcsmatel}. As said above, we have adopted a finite-range interaction, and this allows us to use it, without any change, in both the HF and the BCS parts of our calculations \cite{dec80}. 
 
Our HFBCS calculations give a description of the nuclear ground state in terms  of the s.p. wave functions. The total energy of the even-even nucleus with $A$ nucleons and $Z$ protons can be expressed as:
\beqn
\label{eq:ehf}
E \, \equiv \, E(A,Z)  &=& \sum_k v^2_k \, \epsilon_k \,  
- \, \frac{1}{2} \, \, \sum_k v^2_k \, \int_0^\infty {\rm d}r_1 \, r_1^2  
\, \left[ {\cal U}_k (r_1)\,+
\,2\, {\cal K}(r_1) \right]\, R^2_k(r_1) \\
\nonumber
&& \hspace*{1.8cm} +\, \half \, \sum_k  v^2_k \, \int_0^\infty {\rm d}r_1 
\, {\rm d}r_2 \,
r_1^2 \, r_2^2 ~ {\cal W}_k(r_1,r_2) \,R_k(r_1)  \,R_k(r_2)\, .
\eeqn

The density distribution of the system does not have any more spherical symmetry, but it depends also on the angular coordinates. 
In order to have an estimate of the non-spherical components of the nuclear density, we expand it in multipoles:
\beq
\rho({\bf r}) \,= \, \sum_k \left| \phi_k(x) \right| ^2 \, = \, \sum_L \rho_L(r) \, Y_{L0}(\Omega) \, ,
\label{eq:sumrhol}
\eeq
where the terms of the density expansion are given by:
\beqn
\nonumber 
\rho_L(r) &=& \int {\rm d} \Omega \, Y_{L0} (\Omega) \, \rho(r,\Omega) 
\\ &=& \frac{1}{\sqrt{4 \pi}} \, \sum_{k} \sum_{\mu_k =- l_k}^{l_k} \sum_{s_k = \pm 1/2} 
v^2_{k} \, (-1)^{\mu_k} \, \hat{l_k}^2\,\hat{j_k}^2\,\hat{L} \, R^2_{k}(r)
 \nonumber \\
&& 
\threej {l_k} {l_k} L 0 0 0 
\threej {l_k} {l_k} L {\mu_k} {-\mu_k} 0
{\threej {l_k} {\half} {j_k} {\mu_k} {s_k} {-m_k}}^2
\, .
\label{eq:rhol}
\eeqn
In this last expression, we have used the Wigner $3j$ symbols \cite{edm57} instead  of the Clebsch-Gordan coefficients. 

We have considered the proton (p) and neutron (n) root mean square (r.m.s.) radii, which summarize the characteristics of the density distributions and are defined as:
\beq
R_\alpha\, \,
= \, \left[
\displaystyle \frac 
{\displaystyle \int {\rm d}^3r \, r^2 \, \rho^{\alpha}(\br)}
{\displaystyle \int {\rm d}^3r \, \rho^{\alpha}(\br)} 
\right]^{\half}
\, =\, \left[
 \displaystyle \frac 
{\displaystyle \int {\rm d}r \, r^4 \, \rho_0^{\alpha}(r)}
{\displaystyle \int {\rm d}r \, r^2 \, \rho_0^{\alpha}(r)}
\right]^\half
 \, , \,\,\,\, \alpha \equiv {\rm p,n} \, .
\label{eq:r.m.s. }
\eeq
Here $\rho_0^\alpha(r)$ indicates the $L=0$ multipole of the proton 
or neutron density, which is calculated by using Eq.~(\ref{eq:rhol}) but restricting the sum on $k$ to proton or neutron s.p. states only.

Nuclear deformations have been estimated by using the parameter 
\beq
\beta_2 \,=\, \sqrt{ \frac{ 5 \pi} {9} } \, \frac{1} {A R^2} \,Q_{20} \, ,
\label{eq:beta2}
\eeq 
which simplifies the comparison between nuclei with different size and number of nucleons. 
In the previous equation, $R= 1.2\, A^{1/3}\,$fm and
\beq
Q_{20} \,= \,\sqrt{ \frac{16 \pi} {5} } \braket{\Phi | r^2 Y_{20} | \Phi}\,
= \, \sqrt{ \frac{16 \pi} {5} }\, \int {\rm d}r\, r^4 \rho_2(r) 
\label{eq:q2}
\eeq 
indicates the quadrupole moment of the density distribution, with $\rho_2(r)$ the $L=2$ term of the nuclear density, defined in Eq.~(\ref{eq:rhol}).

We have also calculated the charge radii, which are given by:
\beq
R_{\rm charge}\, \,
= \, \left[
\displaystyle \frac 
{\displaystyle \int {\rm d}r \, r^4 \, \rho_{\rm charge}(r)}
{\displaystyle \int {\rm d}r \, r^2 \, \rho_{\rm charge}(r)} 
\right]^{\half}
 \, .
\label{eq:rcharge}
\eeq
The charge distribution, $\rho_{\rm charge}(r)$, is obtained by folding the point-like proton density, $\rho^{\rm p}(r)$, with the charge proton form factor. We have used a dipole parameterisation of this form factor \cite{pov93}, having verified that other, more accurate, expressions produce differences smaller than the numerical accuracy of our calculations. 

%
\section{Details of the calculations} 
\label{sec:calculations} 

The only physics input of our calculations is the effective nucleon-nucleon interaction. We have used a finite-range interaction of Gogny type, specifically the D1S parameterization \cite{ber91}. In addition, we have considered the interaction D1ST2a \cite{ang12}, obtained by adding to the D1S force a tensor part of the form:
\beq
V_{\rm tensor} (\br_1,\br_2) = 
\left[ V_T \,+\, V_{T \tau}\, \btau(1) \cdot \btau(2) \right] \, S_{12} \, 
\exp \left[ -\,\frac{(\br_1 \,-\, \br_2)^2}{\mu^2_T} \right] \, .
\label{eq:vtensor}
\eeq
In the above expression, $\btau(i)$ indicates the Pauli operator acting on the isospin of the $i$-th nucleon, and $S_{12}$ the usual tensor operator (see Eq.(\ref{eq:tens})).

The values of the parameters of the D1ST2a interaction are those of the D1S force in the common channels. For the parameters of the tensor part of 
D1ST2a, the values $V_T=-77.5\,$MeV,  $V_{T \tau}=57.5\,$MeV and $\mu_T=1.2\,$fm, have been chosen to reproduce the experimental energy splitting between the $1f$ s.p. levels of the $^{48}$Ca nucleus, in a HF calculation, and the empirical excitation energy of the first $0^-$ state of the $^{16}$O nucleus, in a Random Phase Approximation calculation \cite{ang12}. 

As indicated in Appendix \ref{sec:hfmatel},  we separate the contribution 
of the two space coordinates $\br_1$ and $\br_2$ by considering the Fourier transform of the effective nucleon-nucleon interaction. The required integrations in both coordinate and momentum spaces are carried out with the Simpson's technique. A good convergence of up to six significant figures is found by using a set of equally spaced points of $0.1\,$fm in $r$ space and of $0.5\,{\rm fm}^{-1}$ in $q$ space and, respectively, upper integration limits of $15\,$fm and $10\,{\rm fm}^{-1}$.

The radial HF differential equations are solved by using the plane wave expansion technique described in detail in Refs.~\cite{gua82a,gua82b}. The 
iterative procedure stops when the total energies of two consecutive solutions differ by less than $\eta = 10^{-6}\,$MeV. We have used this convergence benchmark in all our calculations. 

In principle, after every iteration where the HF equations (\ref{eq:hfradial}) are solved, the s.p. wave functions just obtained are used in BCS equations in order to modify their occupation probabilities. In practice, we have activated the BCS calculation only when the difference between the total energies of two consecutive HF solutions differ by less than a factor $f\eta$. We have found that, in practice, values of $f$ between 500 and 1000 allow us to obtain a stable convergence of the solutions of the whole problem.

The iterative procedure starts by using the s.p. wave functions obtained by solving the Schr\"odinger equation for a deformed Woods-Saxon potential:
\beq
V_{\rm Woods-Saxon}(r,\Omega) \,=\, 
\frac{U_0}{1+\exp(u)} \,+\, 
\frac {U_{so}}{r} \,\frac{\exp(u)}{[1+\exp(u)]^2} \, {\bf l} \cdot {\bf s}\, + \, V_{\rm C} \, 
-\, \Lambda \, Y_{20}(\Omega) \, ,
\eeq 
where $u=(r-R_0)/a$, ${\bf l}$ and ${\bf s}$ indicate the s.p. orbital angular momentum and spin operators, respectively, and $V_{\rm C}$ the Coulomb potential. Even though the final result of the iterative procedure is independent of the starting set of s.p. wave functions, an appropriate choice of the values of the parameters $U_0$, $U_{so}$, $a$, $R_0$ is crucial to speed up the  convergence. In our calculations, we have used the values indicated in Ref. \cite{volya}, where they have been chosen to reproduce s.p. energies of the odd-even neighbouring nuclei. 

Pragmatically, we have found that values of $|\Lambda|$ of the order of tens of MeV are needed to produce deformed solution. Specifically, we have always used $\Lambda = \pm \,30$ MeV. In all the calculations we have carried out, we observed that our procedure finds an energy minimum with the same type of deformation as that of the starting set of s.p. wave functions. For example, when we started with a prolate deformation, by setting $\Lambda > 0$, we found an energy minimum that maintained the prolate deformation, and viceversa. For this reason, in Sec. \ref{sec:results}, sometimes, we show results obtained for both types of deformations. 

Since the energies of the oblate and prolate solutions obtained for a given nucleus are different, we named \textsl{optimal solution} that with the smaller value of $E(A,Z)$, in other words, the solution providing more binding to the system.

The relevance of the use of the new set of deformed s.p. wave functions has been studied by comparing the HFBCS results with those calculated within the HF+BCS approach of Ref.~\cite{ang14}, where a spherical approximation is adopted. In this latter model, each s.p. state of angular momentum $j$ is $2j+1$ times degenerated, the occupation of each s.p. state is equally distributed on all the possible $z$-axis projections and the full system conserves a spherical symmetry. For this reason we have indicated as 
\textsl{spherical} the results of the HF+BCS calculations. This is the most important difference, from the physics point of view, between the HFBCS and the HF+BCS models.

A second, and more technical, difference between the two approaches is in the treatment of the pairing. In HF+BCS we first carry out a HF calculation and, afterwards, we use the obtained s.p. wave functions to perform a BCS calculation. In HFBCS, the BCS calculations are inserted in the global iterative minimization procedure, connected to the solution of the HF equations.

We have also compared our HFBCS results with those of the deformed HFB calculations performed within a triaxial basis \cite{rin80,egi95,ang01a}. In these latter calculations, the solutions are expanded on a orthonormal basis of harmonic oscillators wave functions with oscillator length $b_0$. The number of harmonic oscillators considered in each quantisation axis, $(n_x,\, n_y, \, n_z)$, must satisfy the following condition for the energy truncation \cite{ang01a}:
\beq
a_x\, n_x \,+\, a_y\, n_y \,+\, a_z \, n_z \, \leq \, N_0 \, .
\eeq
Here $a_x= (qp)^{1/3}$, $a_y=q^{1/3} p^{-2/3}$ and $a_z = p^{1/3} q^{-2/3}$, where $p={R_y}/{R_x}$ and $q={R_z}/{R_x}$ are ratios between the semi-axes of the matter distribution. The results we have presented 
here have been obtained with $N_0=9$, and by using  $b_0=1.01\,A^{1/6}\,$fm as indicated in Ref.~\cite{rin80}.
 
By adequately choosing the values of $p$ and $q$, a deformation shape can 
be selected. In our calculations we have used $p=1$, that imposes the symmetry around the $z$ axis, and $q=1.3$ or $q=0.7$ to select either a prolate or an oblate initial deformation, respectively. As it occurred in the HFBCS calculations, the deformation initially selected is maintained in the final solution provided by the HFB calculations. 

We have tested the validity of our HFB results by verifying that the values of the total energies and density r.m.s. radii coincide with those presented in the compilation of Bruy\`eres \cite{cea,capote09}. 
 
We close this section by mentioning what is known in the literature as the {\sl neutron gas problem} \cite{dob84}. The BCS calculations allow the presence of a long unphysical tail of the nuclear density distributions due to the contributions of slightly bound nucleons. We have shown in Ref. 
\cite{ang19} that this is only a formal problem, since in actual calculations the numerical impact of these tails is irrelevant.

\section{Results} 
\label{sec:results} 

In this section we present some selected results of our calculations with 
the goal of pointing out the combined effects of  deformation and tensor  
force. We have considered 16 even-even $Z=34$ isotopes, from $^{64}$Se to $^{94}$Se, and 10 even-even  $N=34$ isotones: $^{52}$Ar, $^{54}$Ca, $^{56}$Ti, $^{58}$Cr, $^{60}$Fe,  $^{62}$Ni, $^{64}$Zn, $^{66}$Ge, $^{68}$Se, and $^{70}$Kr.

\subsection{Comparison with HFB and HFBCS calculations}

In the first step of our study, we have tested the reliability of our calculations. For this purpose, we have compared our HFBCS results with those of well established nuclear models. In our case, we refer to the deformed HFB results of the Bruy\`eres compilation \cite{cea,capote09} and to those we have obtained by using the HFB approach of Ref.~\cite{ang01a} described in Sect.~\ref{sec:calculations}.

\begin{figure}[!ht] 
\begin{center} 
\includegraphics [width=7.5cm,angle=0]{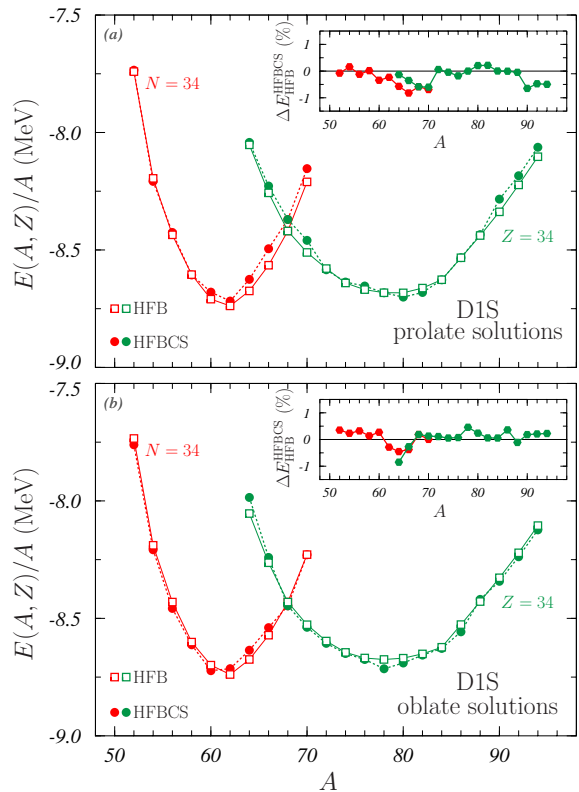} 
\vskip -0.3 cm 
\caption{\small 
Total energies per nucleon of the nuclei considered calculated with the 
D1S interaction. Our HFBCS results (full circles) are compared to those of the HFB approach of Ref. \cite{ang01a}. The energies obtained for the prolate and oblate solutions are shown in the upper and lower panels, respectively. In the insets, we show the relative differences $\Delta E^{\rm HFBCS}_{\rm HFB}$, in percentage, as defined in Eq. (\ref{eq:Breldif}). We use the red color to indicate the results of the $N=34$ isotones and the green color for those of the Se, $Z=34$, isotopes. The lines are drawn to guide the eyes. 
}
\label{fig:D1Stest} 
\end{center} 
\end{figure} 

The total energies per nucleon, $E/A$, with $E$ given by Eq.~(\ref{eq:ehf}), of the nuclei considered are shown in Fig.~\ref{fig:D1Stest}. Our HFBCS results are indicated by the full circles and the benchmark HFB results by the empty squares. Since, as pointed out in Sect. \ref{sec:calculations}, in both calculations the deformation initially selected is maintained in the iterative procedure up to the final solution, we show separately the results obtained for prolate (upper panel) and oblate (lower panel) 
deformations. The differences between the two calculations are emphasized 
in the insets where the relative differences
\beq
\Delta E^{\rm HFBCS}_{\rm HFB} \, = \, \frac{(E/A)_{\rm HFBCS} \,- \, (E/A)_{\rm HFB}}{(E/A)_{\rm HFB}} \, ,
\label{eq:Breldif}
\eeq
between the corresponding results are shown. 

The comparison between the energies obtained with the two nuclear models is very satisfactory. The relative differences $\Delta E^{\rm HFBCS}_{\rm 
HFB}$ are smaller than 1\% for both type of deformations. We observe that 
most of the HFB energies in the prolate solutions (upper panel) are smaller than those of the HFBCS calculations. This is clearly indicated by the 
negative values of relative differences shown in the inset. The oblate solutions (lower panel) exhibit the opposite behavior. 

\begin{figure}[!t]
\hspace*{0.2cm}
\begin{minipage}[c]{0.45\linewidth}
\includegraphics[width=6.5cm,angle=0]{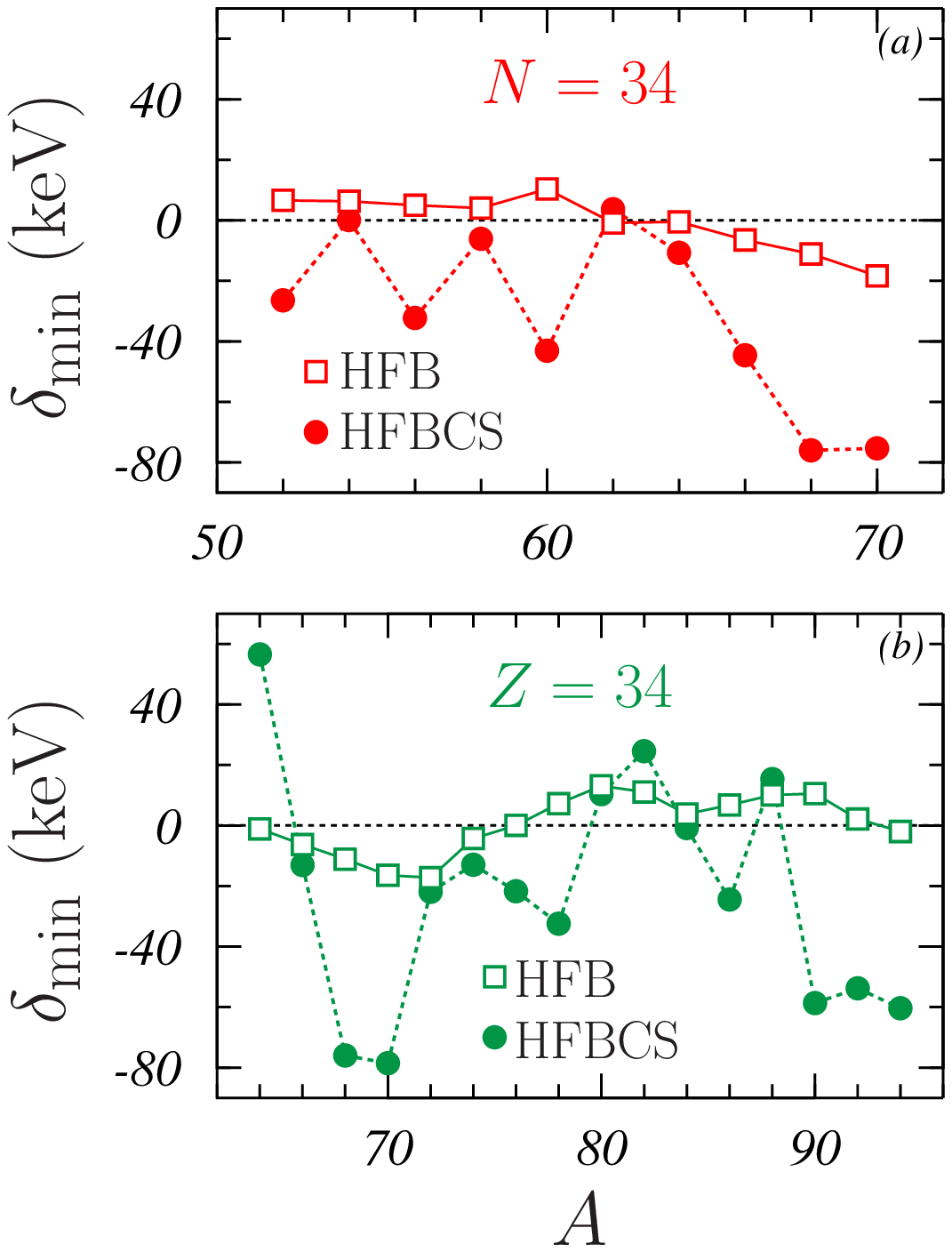} 
\vspace{-0.3cm}
\caption{\small 
Values of $\delta_{\rm min}$, defined in Eq. (\ref{eq:delta-min}), obtained in  
HFBCS (full circles) and HFB (empty squares) calculations carried out with the D1S interaction. 
The results for the $N=34$ isotones are shown in panel (a)  and those of the Se isotopes in panel (b). }
\label{fig:DEFmin}  
\end{minipage}
\hspace*{0.5cm}
\begin{minipage}[c]{0.45\linewidth}
\vspace*{0.75cm} 
\includegraphics[width=6.5cm,angle=0]{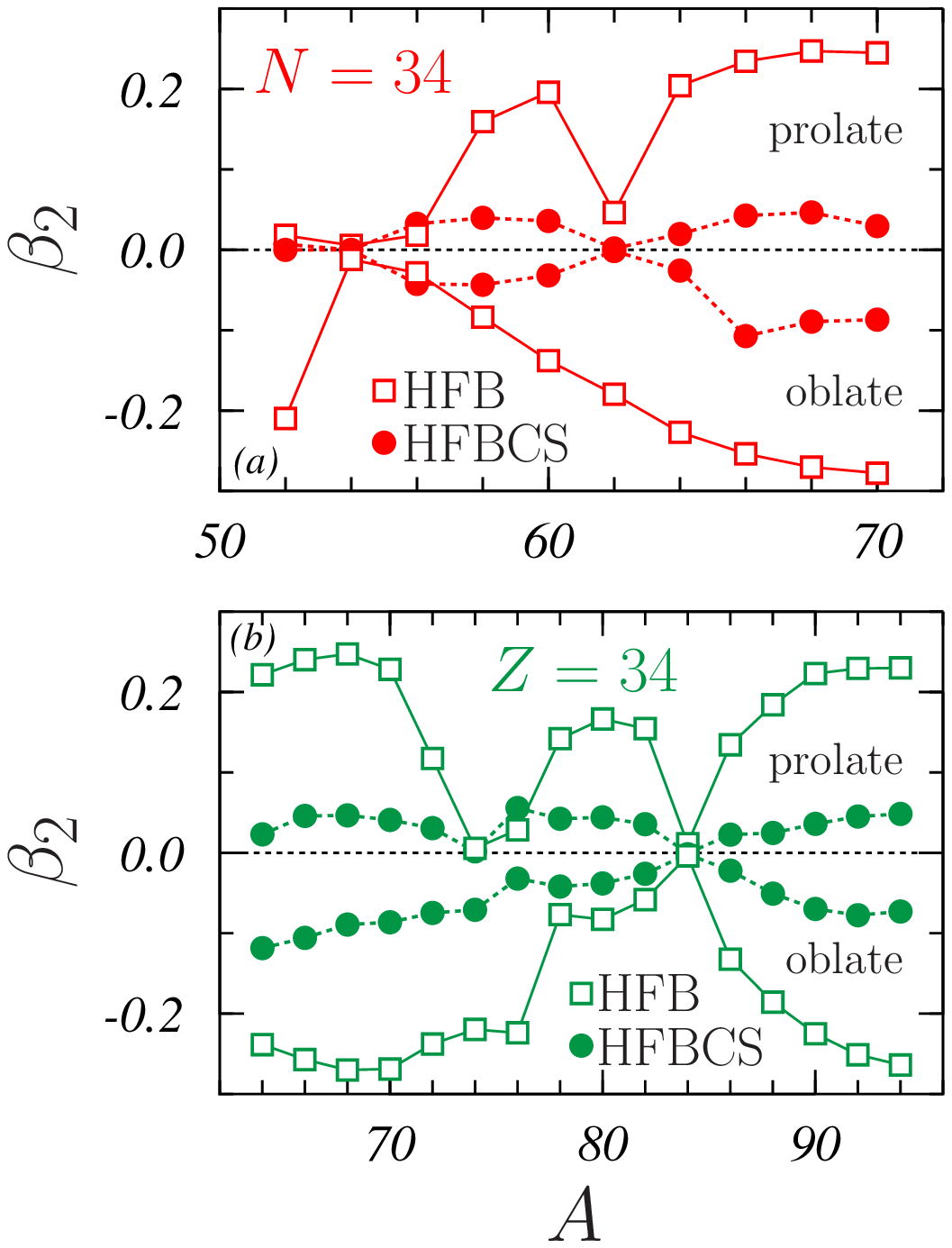} 
\vspace{-0.3cm}
\caption{\small 
Deformation parameter $\beta_2$, defined in Eq. (\ref{eq:beta2}), obtained for (a) the $N=34$ isotones, and (b) the Se isotopes, considered. The 
full circles indicate the HFBCS results, and the empty squares those found in the HFB calculations, all of them carried out with the D1S interaction. The values found for both prolate ($\beta_2 > 0$) and oblate ($\beta_2 < 0$) solutions are shown.
}
\label{fig:DEFtest}
\end{minipage}
\end{figure}

We show in Fig.~\ref{fig:DEFmin} the differences between the absolute values of the total energies per nucleon of the prolate and oblate solutions:
\beq
\delta_{\rm min} \, = \, |(E/A)_{\rm prolate}|\, -\, |(E/A)_{\rm oblate}| \, .
\label{eq:delta-min}
\eeq
Positive (negative) values of $\delta_{\rm min}$ indicate that the {\sl optimal solution} has prolate (oblate) deformation.

For HFBCS (full circles), these differences are, at most, $80\,$keV, in absolute value, and for HFB (empty squares) they are even smaller. These numbers are close to the numerical uncertainty of the calculations, which is of a few tens of keV. This fact has been named \textsl{shape coexistence} \cite{lal98}. 
It means that, in practice, the {\sl optimal solution}, and consequently the nuclear shape, is not so well defined in many cases.

According to these results, mean-field like approaches might not be enough to have a good description of these nuclei and dynamical, beyond mean-field correlations could be important. Within this context, angular momentum and particle number projections and fluctuations of the collective deformation
parameters might play a relevant role \cite{rod00,ang01bb,rod10}.

The results of Fig.~\ref{fig:DEFmin} also indicate that, often, the deformation of the {\sl optimal solution} is not the same in HFB and HFBCS calculations. This happens for all the nuclei up to $A=62$ in the $N=34$ 
chain, and for quite a few Se isotopes. We have found that, in some cases, this also occurs when we compare our HFB results and those of Bruy\`eres \cite{cea}. 

It is worth pointing out that the total energies of the {\sl optimal solutions} obtained in HFBCS are slightly lower (0.5\% at most) than those of 
the HFB for all cases considered, with the exception of the five nuclei with $A=62-66$.

If the agreement between the total energies obtained with HFBCS and HFB is satisfactory, the situation is quite different when the  deformation parameters are considered. We show in Fig.~\ref{fig:DEFtest} the $\beta_2$ values obtained in HFBCS (full circles) and HFB (empty squares) calculations. Positive and negative values of $\beta_2$ refer to prolate and oblate deformations, respectively. The values of the $\beta_2$ of the HFBCS solutions are always remarkably smaller, in absolute value, than those found with HFB, although their overall behavior, as $A$ varies, is similar in 
both calculations. The more relevant exception is that of the $^{62}$Ni which is spherical in HFBCS and deformed in HFB, especially in the case of 
the oblate solution. On the other hand, the $^{54}$Ca and $^{84}$Se nuclei are clearly spherical in both approaches. 

Finally, the effects of the deformation have been evaluated by comparing the total energies obtained in the HFBCS calculations with those calculated within the HF+BCS approach of Ref.~\cite{ang14}, where a spherical approximation is adopted, as we have briefly discussed in Sect. \ref{sec:calculations}. The results of the two types of calculations are very similar, the relative differences being $\pm 1.5\%$ at most. In general, the deformed HFBCS produces {\sl optimal solutions} that are more bound than the 
spherical ones. We have found only three exceptions:  $^{54}$Ca, $^{62}$Ni and $^{64}$Zn. 

\subsection{Tensor effects.}
\label{sec:tensor}

\begin{figure}[!t] 
\begin{center} 
\includegraphics [width=7.5cm,angle=0]{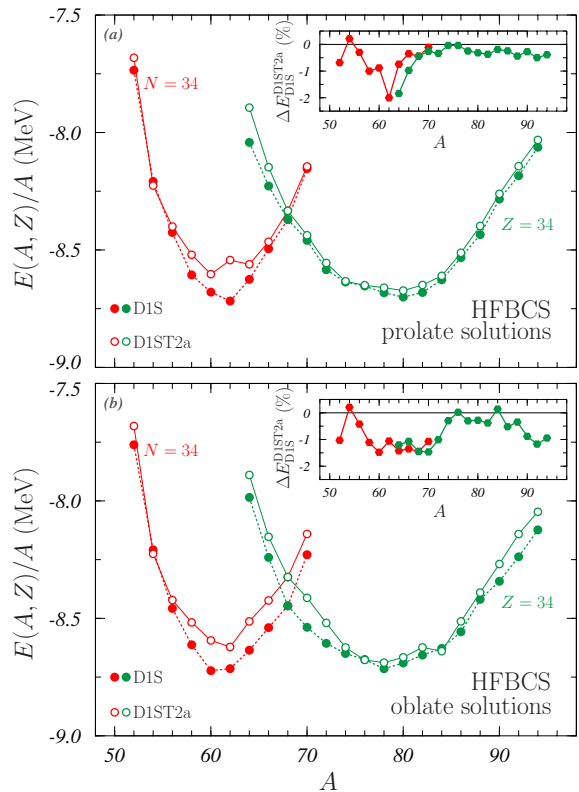} 
\vskip -0.3 cm 
\caption{\small 
Total energies per nucleon obtained within the HFBCS approach by using the D1S (full circles) and D1ST2a (empty circles). In the upper and lower panels we show the results of the prolate and oblate solutions, respectively. Red symbols indicate the results corresponding to the $N=34$ isotones, while green symbols show those of the Se nuclei. In the insets we give the relative differences between D1ST2a and D1S results, in percentage, 
calculated as indicated by Eq.~(\ref{eq:B2reldif}). \\
}
\label{fig:tensor-bin} 
\end{center} 
\end{figure}

The impact of the tensor force on the HFBCS total energies is shown in Fig~\ref{fig:tensor-bin}, where we compare the results obtained with the D1ST2a (empty circles) and the D1S (full circles) forces.  Also in this case, we present separately the $E/A$ values corresponding to the prolate (upper panel) and oblate (lower panel) deformations. We show in the insets the relative differences between the results obtained with the two interactions:
\beq
\Delta E^{\rm D1ST2a}_{\rm D1S} \, = \, \frac{(E/A)_{\rm D1ST2a} \,- \, 
(E/A)_{\rm D1S}}{(E/A)_{\rm D1S}} \, .
\label{eq:B2reldif}
\eeq

In general, the tensor interaction generates less binding. The only exceptions we have found are those of the $^{54}$Ca, for both deformations, and of the $^{84}$Se, for the oblate solution. On the other hand, and in agreement with our previous studies \cite{ang12}, 
we notice that the role of the tensor force on this observable is rather small. We identify values of the relative differences which are 2\% at most. These differences are, in any case, larger than those between the HFBCS and HFB energies obtained with the D1S interaction (see Fig.~\ref{fig:D1Stest}). 

\begin{figure}[!t]
\hspace*{0.2cm}
\begin{minipage}[c]{0.45\linewidth}
\includegraphics[width=6.5cm,angle=0]{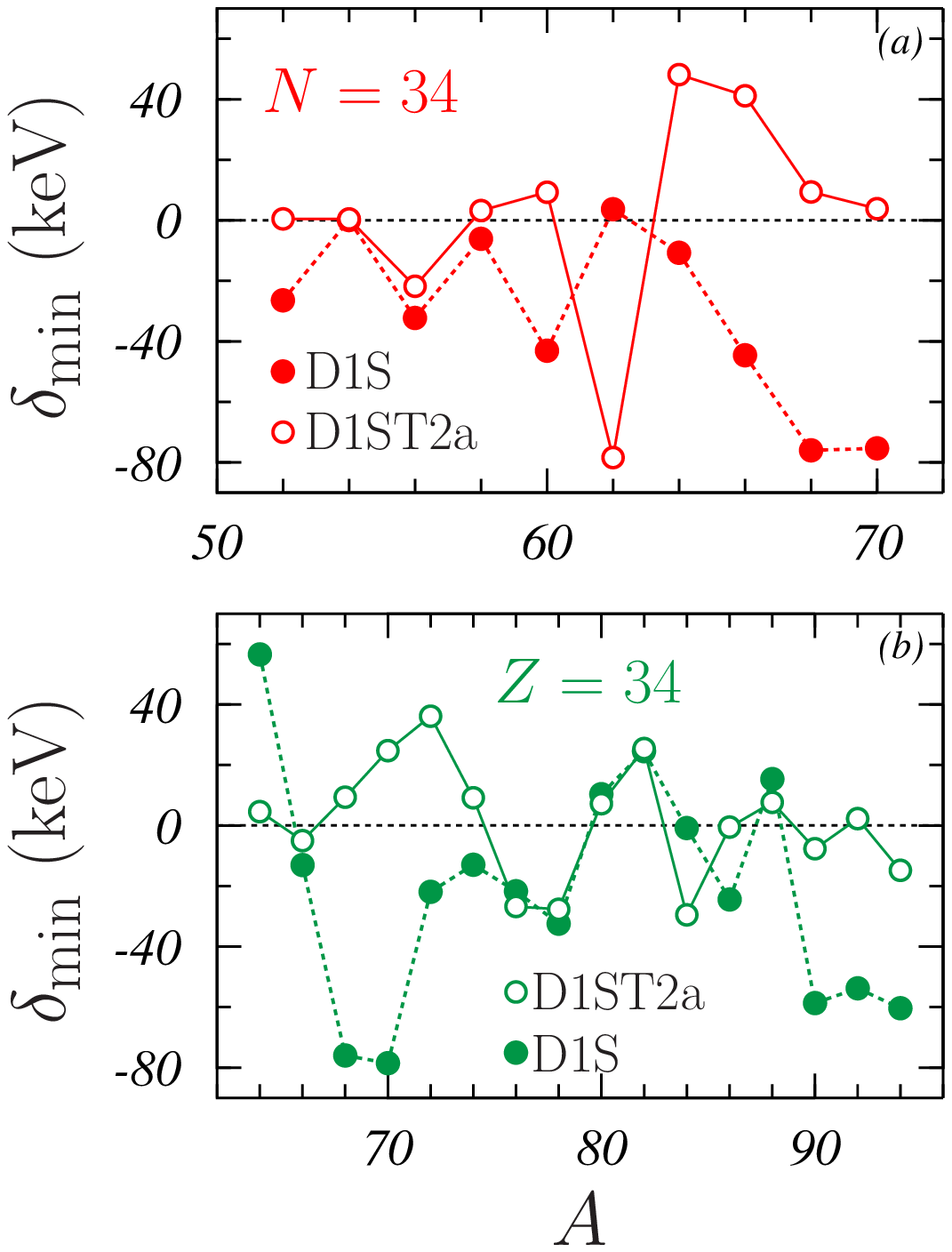}  
\vspace{-0.3cm}
\caption{\small 
Values of $\delta_{\rm min}$, defined in Eq. (\ref{eq:delta-min}), obtained in HFBCS calculations with the D1S (full circles) and the D1ST2a (empty circles) interactions. The results for the $N=34$ isotones are shown in panel (a)  and those of the Se in panel (b).  }
\label{fig:tens-edif} 
\end{minipage}
\hspace*{0.5cm}
\begin{minipage}[c]{0.45\linewidth}
\vspace*{0.75cm} 
\includegraphics[width=6.5cm,angle=0]{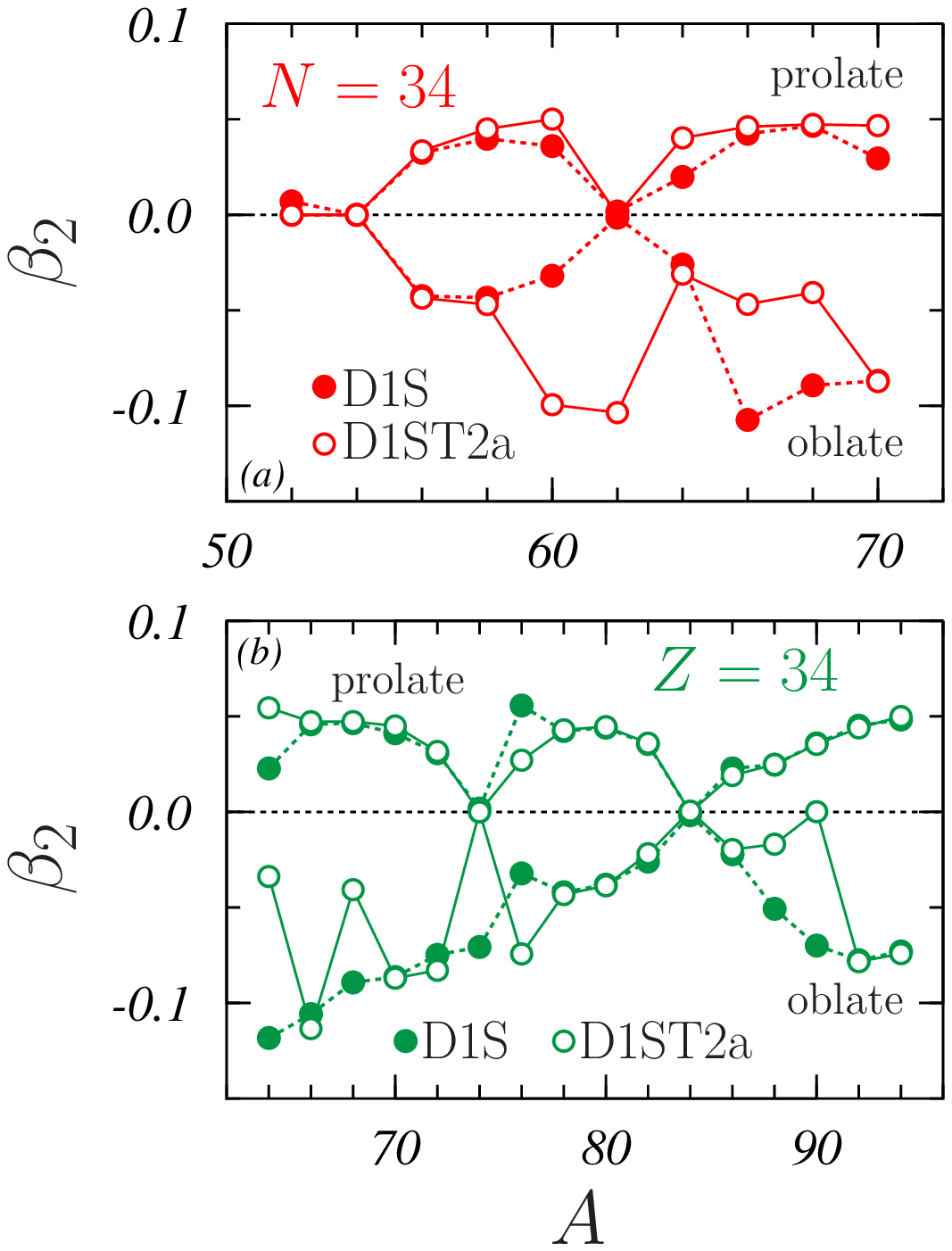} 
\vspace{-0.3cm}
\caption{\small 
Values of the deformation parameter $\beta_2$, defined in Eq.~(\ref{eq:beta2}), calculated with our HFBCS approach for (a) the $N=34$ and (b) the $Z=34$ chains and for both the prolate ($\beta_2 > 0$) and oblate ($\beta_2 < 0$) solutions. The results obtained with the D1S interaction are 
shown by the full circles and those found with the D1ST2a force by the empty circles.
}
\label{fig:tens-bdif} 
\end{minipage}
\end{figure}

In Fig.~\ref{fig:tens-edif} we show the differences $\delta_{\rm min}$, defined in Eq.~(\ref{eq:delta-min}), calculated with the D1ST2a (empty circles) and the D1S (full circles) forces. As already pointed out, a change 
in the sign of $\delta_{\rm min}$ implies a change of the deformation of the {\sl optimal solution} and one can see that this occurs in about half 
of the nuclei investigated.

The differences $|\delta_{\rm min}|$ between prolate and oblate energies are, in average, smaller when the tensor interaction is considered. We have obtained an average value of about 17 keV for D1ST2a against about 34 keV for the D1S. In the case of the HFB calculations with the D1S force (empty squares in Fig.~\ref{fig:DEFmin}), the average value found is $\sim 
7\,$keV.

The effects of tensor force on the deformation are shown in Fig.~\ref{fig:tens-bdif}, where the values of $\beta_2$ calculated with the D1S (full circles) and D1ST2a (empty circles) interactions are shown. Also in this case, we show separately the results obtained for prolate and oblate deformations. 

The values of $\beta_2$ for the prolate deformation are almost the same for the two interactions. In the case of the oblate solutions, we have found noticeable differences in many of the nuclei investigated, the most noteworthy are those of the $^{62}$Ni nucleus, which loses its spherical shape when the D1ST2a force is used, and of the $^{74}$Se and $^{90}$Se nuclei, which, on the contrary, become spherical. 

\begin{table} [!bh]
\begin{center}
\begin{tabular} {ccc c ccc} 
\hline\hline
\multicolumn{3}{c}{$N=34$} &~~~&  \multicolumn{3}{c}{$Z=34$} \\ \cline{1-3} \cline{5-7}
nucleus & D1S & D1ST2a && nucleus & D1S & D1ST2a \\ 
\hline
\rule{0cm}{0.3cm}
$^{52}$Ar & $ 0.000$ & $ 0.000$ && $^{64}$Se & $ 0.023$ & $ 0.054$ \\
$^{54}$Ca & $ 0.000$ & $ 0.000$ && $^{66}$Se & $-0.106$ & $-0.113$ \\
$^{56}$Ti & $-0.042$ & $-0.044$ && $^{68}$Se & $-0.089$ & $ 0.047$ \\
$^{58}$Cr & $-0.043$ & $ 0.045$ && $^{70}$Se & $-0.086$ & $ 0.045$ \\
$^{60}$Fe & $-0.032$ & $ 0.050$ && $^{72}$Se & $-0.075$ & $ 0.032$ \\
$^{62}$Ni & $ 0.002$ & $-0.104$ && $^{74}$Se & $-0.071$ & $ 0.000$ \\
$^{64}$Zn & $-0.026$ & $ 0.040$ && $^{76}$Se & $-0.032$ & $-0.074$ \\
$^{66}$Ge & $-0.107$ & $ 0.046$ && $^{78}$Se & $-0.042$ & $-0.043$ \\
$^{68}$Se & $-0.089$ & $ 0.047$ && $^{80}$Se & $ 0.044$ & $ 0.045$ \\
$^{70}$Kr & $-0.087$ & $ 0.047$ && $^{82}$Se & $ 0.036$ & $ 0.036$ \\
          &          &          && $^{84}$Se & $-0.002$ & $ 0.000$ \\
          &          &          && $^{86}$Se & $-0.022$ & $-0.020$ \\
          &          &          && $^{88}$Se & $ 0.025$ & $ 0.025$ \\
          &          &          && $^{90}$Se & $-0.070$ & $ 0.000$ \\
          &          &          && $^{92}$Se & $-0.078$ & $ 0.044$ \\
          &          &          && $^{94}$Se & $-0.073$ & $-0.074$ \\
\hline\hline
\end{tabular}
\caption{Values of the deformation parameter $\beta_2$ 
for the {\sl optimal solutions} of the nuclei studied in the present work.}
\label{tab:gs-def}
\end{center}
\end{table}

\newpage \clearpage

Up to now, we have presented our results separately for both prolate and oblate deformations. For the comparison with the experimental data we consider only the results of the {\sl optimal solutions}. We show in Table~\ref{tab:gs-def} the corresponding $\beta_2$ values calculated with the D1S and D1ST2a interactions. We remark that, quite often, the type of the deformations of the {\sl optimal solutions} for the two interactions are different. Out of the 26 nuclei considered, only eight, two of them are spherical, maintain the same type of deformation when the tensor terms  are included in the interaction.

\begin{figure}[!h] 
\begin{center} 
\includegraphics [width=7.5cm,angle=0]{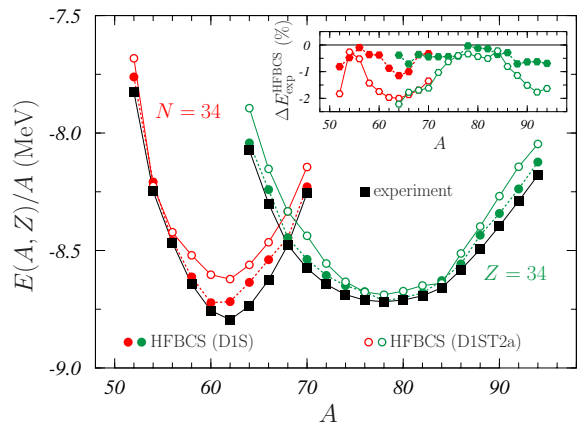} 
\vskip -0.3 cm 
\caption{\small 
Total energies per nucleon of the HFBCS {\sl optimal solutions} obtained with the D1S (full circles) ad D1ST2a (empty circles) interactions, compared to the experimental values of Ref. \cite{bnlw} (full squares). 
The insets show the relative differences between theory and experiment, as defined in Eq. (\ref{eq:Bexp-reldif}), for the D1S and the D1ST2a forces, full and empty symbols respectively.
}
\label{fig:dbexp} 
\end{center} 
\end{figure}

The total energies per nucleon corresponding to the {\sl optimal solutions} obtained in  HFBCS with the D1S (full circles) and D1ST2a (empty circles) interactions, are compared in Fig.~\ref{fig:dbexp} with the experimental data of Ref.~\cite{bnlw} (full black squares). We show in the inset the relative differences between our results and the experimental values:
\beq
\Delta E^{\rm HFBCS}_{\rm exp} \, 
= \, \frac{(E/A)_{\rm HFBCS} \,- \, (E/A)_{\rm exp}}{(E/A)_{\rm exp}} \, .
\label{eq:Bexp-reldif}
\eeq

The differences with the experimental energies are quite small: they are, 
at most, 1\% for the D1S interaction and reach values of about $2\%$ for the D1ST2a. The HFBCS {\sl optimal solutions} obtained with both interactions are less bound than the experimental values. From the point of view of the variational principle, this is quite reassuring. 

The effect of the tensor force consists of a reduction of the nuclear binding. This has been already indicated for prolate and oblate solutions separately, and Fig.~\ref{fig:dbexp} shows that it is also true for the {\sl optimal solutions}. A critical discussion of these results is in order, 
since the better agreement of the D1S results with the experimental data can be misinterpreted. The values of the parameters of the D1S interactions have been optimised to reproduce experimental binding energies and charge r.m.s. radii \cite{ber91,cha07t}. As said above, the tensor terms  in D1ST2a have been added without changing the other parameters of the force. For this reason, it is plausible that the the inclusion of a new term in the interaction is worsening the overall agreement with the experiment. 

In the Se isotope chain, the nuclei with the largest binding energy per nucleon are the $^{78}$Se and $^{80}$Se. Our results indicate that $^{78}$Se is more bound than $^{80}$Se in agreement with the experiment. In fact, $^{78}$Se is the nucleus showing the smallest difference between our HFBCS (D1S) results and the experiment. In the $N=34$ isotones the experimental $|E(A,Z)/A|$ value is slightly smaller in $^{60}$Fe than in  $^{62}$Ni while the opposite occurs in the HFBCS calculation.

\begin{figure}[!bh] 
\begin{center} 
\includegraphics [width=11cm,angle=0]{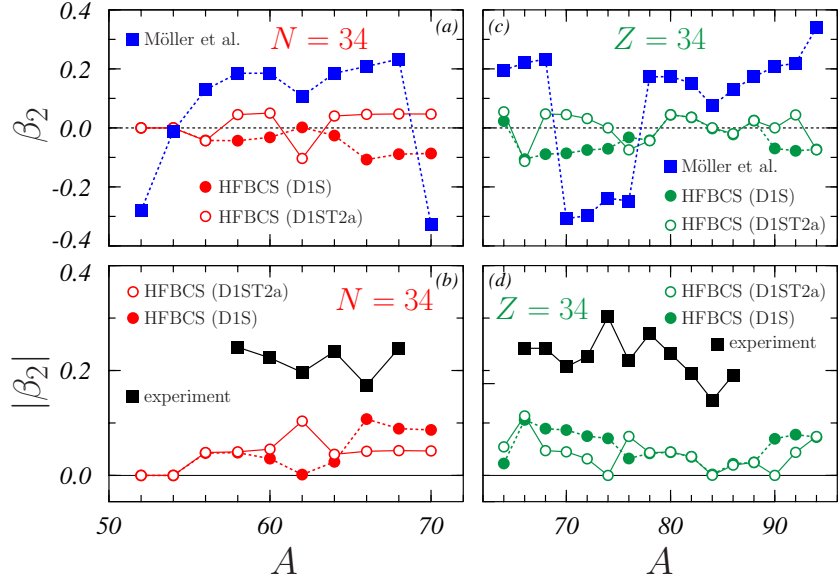} 
\vskip -0.3 cm 
\caption{\small 
Upper panels: deformation parameters $\beta_2$ of the HFBCS {\sl optimal solutions} obtained with the D1S (full circles) and D1ST2a (empty circles) interactions compared to the values tabulated by M\"oller {\it et al.} \cite{mol16} (blue full squares). Lower panels: absolute values of the $\beta_2$ obtained for the HFBCS {\sl optimal solutions} compared to the available experimental values of Ref. \cite{bnlw}  (black full squares). 
}
\label{fig:betaexp} 
\end{center} 
\end{figure} 

\begin{figure}[!bh] 
\begin{center} 
\includegraphics [width=6.5cm,angle=0]{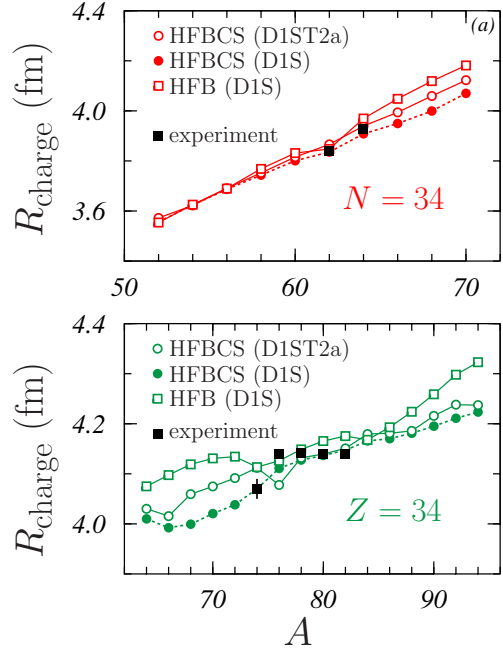}  
\vskip -0.3 cm 
\caption{\small 
Charge radii for the nuclei considered calculated according to Eq.~(\ref{eq:rcharge}). The values of the HFBCS {\sl optimal solutions} obtained with the D1ST2a interaction (empty circles) and with the D1S interaction (full circles) are compared to the values found in HFB calculations \cite{ang01a} by using the D1S force (empty squares) and 
to the experimental values of Ref. \cite{ang13} (full black squares). Results for the $N=34$ isotones are shown in panel (a) and those for the Se isotopes in panel (b). 
}
\label{fig:charge-rad} 
\end{center} 
\end{figure} 

We have investigated the two-neutron separation energies for the $N=34$ 
isotones and the two-proton separation energies for the Se isotopes. We have not found relevant effects due to the presence of the tensor force. On the other hand, the agreement with the experiment is good. 

We present in Fig. \ref{fig:betaexp} the values of $\beta_2$ of the {\sl optimal solutions} obtained in the HFBCS calculations by using the D1S (full circles) and D1ST2a (empty circles) forces, and we compare them with some empirical estimations. In the upper panels the comparison is done with the values of the semi-empirical model of M\"oller et al. \cite{mol16} (blue full squares). Even though no specific patterns are observed, our deformations are smaller, in absolute value, than those of Ref.~\cite{mol16}. Quite often, the deformation of the {\sl optimal solution} obtained with the D1ST2a interaction has different sign than that of the D1S interaction, but also in this case we cannot identify any trend. 

In the lower panels of  Fig. \ref{fig:betaexp} our HFBCS results are compared with the empirical data of Ref.~\cite{bnlw} (full black squares). In this case, we have considered $|\beta_2|$, since the sign of the deformation in these empirical data is usually undetermined. Again, it is evident 
that our approach generates smaller $|\beta_2|$ values than the experimental ones. However, also this comparison should be considered with caution. The $\beta_2$ values of Ref.~\cite{bnlw} have been obtained by assuming that the first $2^+$ excited state is due to a rotation of the deformed nucleus described with a sempi-empirical liquid drop model. The assumptions of this procedure are quite strong, and they lead to assign ground state deformations ground state deformations even to nuclei that are well know to be spherical: for example, 
$\beta_2=0.353$ is quoted for $^{16}$O due to the presence of a $2^+$ state at $6.917\,$MeV.

\subsection{Density distributions and  r.m.s. radii}
\label{sec:r.m.s. }

In this section, we present the results of our HFBCS model concerning proton, neutron and charge density distributions, and their r.m.s. radii. 

We have tested the reliability of our study by comparing our HFBCS results with those of HFB calculations carried out with the D1S interaction. The relative differences between the corresponding r.m.s. radii of the proton and neutron density distributions are smaller than 3\% and 1.5\% for the oblate and prolate solutions, respectively. On the other hand, the relative differences obtained with the D1S and D1ST2a interactions in the HFBCS approach are even smaller: at most 1.1\% for the prolate solutions and 1.7\% for the oblate ones. Contrary to what we have found for the total 
energies, in this case, the effect of the tensor is smaller than the differences between HFB and HFBCS results. 

In order to study the effect of the deformation, we have calculated the  r.m.s. radii with the HF+BCS approach. We have found that  the largest relative difference with the HFBCS r.m.s radii of the {\sl optimal solutions} is about 1\%.

The situation is well summarized in Fig.~\ref{fig:charge-rad}, where we show the r.m.s. charge radii of the HFBCS {\sl optimal solutions} of the nuclei studied, calculated, according to Eq.~(\ref{eq:rcharge}), with both 
the D1S (full circles) and D1ST2a (empty circles) interactions. These results are compared to the experimental values taken from the compilation of Ref.~\cite{ang13} (black full squares) and to those obtained in HFB with the D1S force (empty squares). These latter charge radii have been calculated by using the proton distributions of Ref.~\cite{capote09}. 

For the D1S interaction, the HFB radii are larger than those of our HFBCS 
calculations by 3\% at most. Also the charge radii obtained with the D1ST2a force in HFBCS are slightly larger than those found with the D1S interaction, but the differences are smaller than 1.5\%.

The comparison with the experimental data is limited to only seven nuclei. These few data are well described by all the three types of calculations, even though, globally, the best agreement is obtained for the HFBCS calculations with the D1S interaction.

\begin{figure}[!b] 
\begin{center} 
\includegraphics [width=11cm,angle=0]{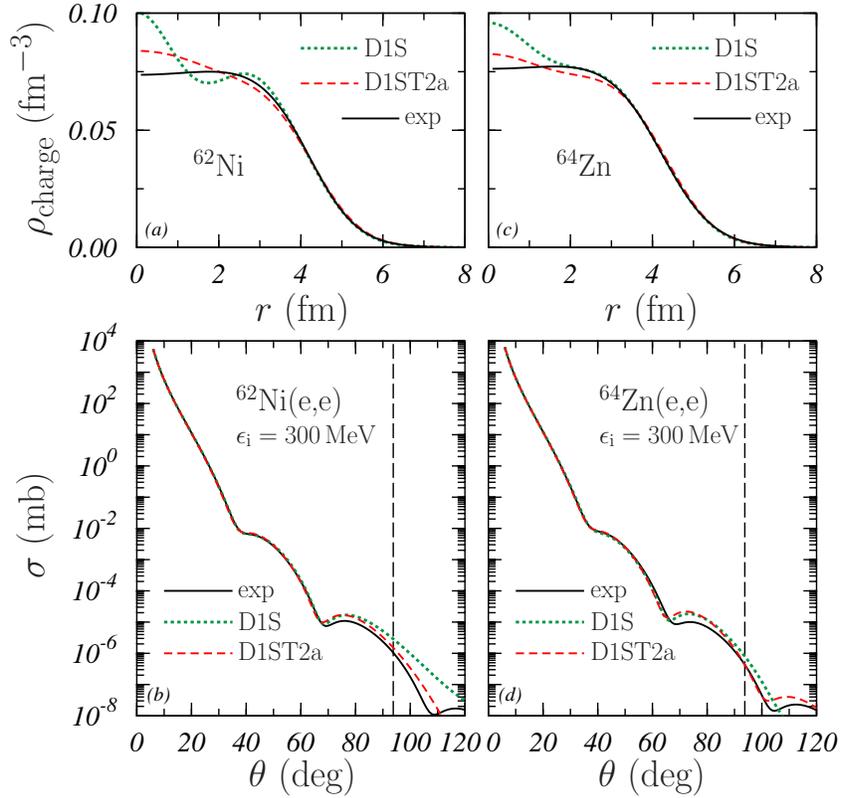} 
\vskip -0.3 cm 
\caption{\small 
Upper panels: charge densities calculated with the D1S (dotted lines) 
and D1ST2a (dashed lines) interactions compared to the empirical densities of Ref.~\cite{dej87} (full lines) for the nuclei $^{62}$Ni, panel (a), and $^{64}$Zn, panel (c). Lower panels: elastic electron scattering cross sections calculated by using the charge densities shown in the upper panels. We have considered an electron initial energy of $300\,$MeV. The vertical dashed lines indicate the scattering angle corresponding to the maximum value of the momentum transfer of the data considered in the fit to the empirical densities.
}
\label{fig:densexp} 
\end{center} 
\end{figure} 

We have analyzed in detail the density distributions and, as example of this study, we show in Fig.~\ref{fig:densexp} the results for
the two nuclei of our set of isotopes and isotones whose charge distributions are available in the compilation of Ref.~\cite{dej87}:
$^{62}$Ni and $^{64}$Zn.

In the upper panels of the figure, we compare the HFBCS charge distributions with the empirical ones. The agreement between them is excellent at the surface, and this explains the good description of the experimental charge radii, which are mostly sensitive to this part of the distributions. 
Remarkable differences are evident in the nuclear interior, the region where correlations of various types, long- and short-ranged, are most effective \cite{ang01}. The oscillations of the distributions obtained with the D1S force are smoothed by the presence of the tensor force, which produces charge densities closer to the empirical ones. 

In order to frame the above discussion in a proper perspective, we remember that the empirical charge densities are tailored to fit elastic electron scattering cross sections. These experimental data have been measured within a restricted range of momentum transfer values, which, in our case, for both nuclei considered, goes up to $q_{\rm max} =2.2\,$fm$^{-1}$ \cite{dej87}. In the lower panels of Fig.~\ref{fig:densexp} we show the elastic 
electron scattering cross sections calculated in Distorted Wave Born Approximation \cite{ann95a} by using the charge densities shown in the upper panels. We have assumed an incident electron energy of $300\,$MeV. In this kinematic conditions the value of $q_{\rm max}$ is reached at $\theta_{\rm max}=93.79\,$deg, which is indicated in the figure by the vertical dashed lines. As a consequence, the comparison between theoretical and empirical cross sections is meaningful only for $\theta <  \theta_{\rm max}$, where a good agreement with the experiment is observed.

\begin{figure}[!bh] 
\begin{center} 
\includegraphics [width=11cm,angle=0]{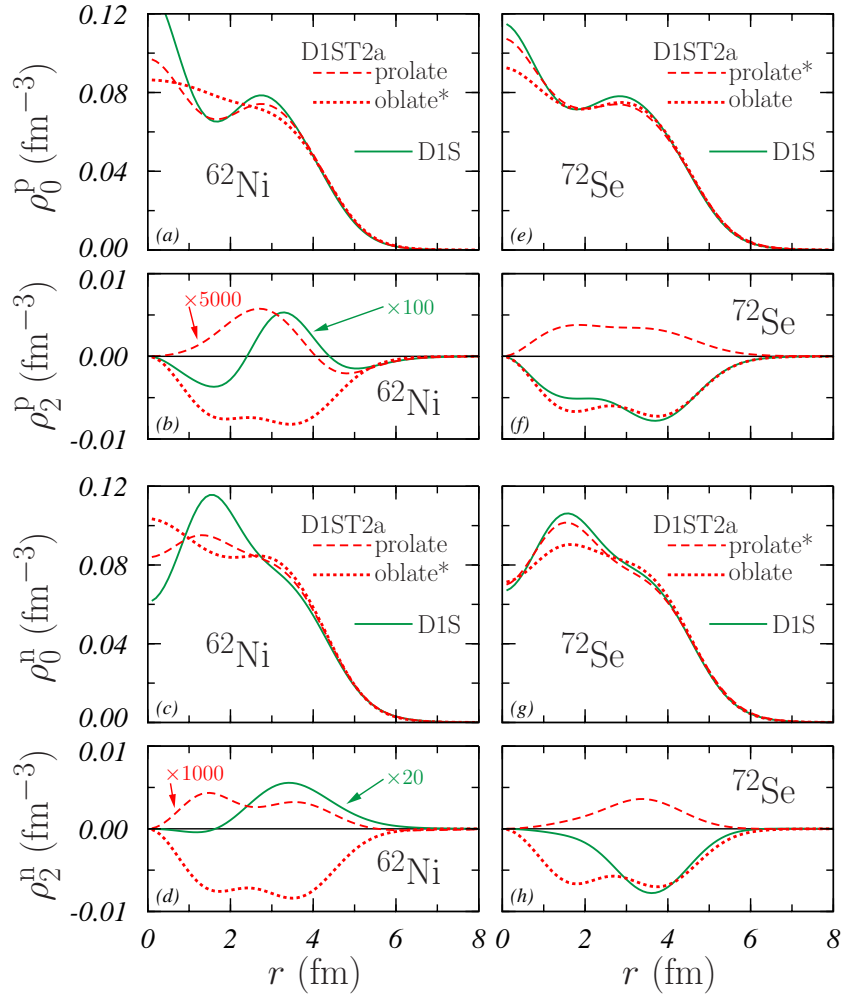} 
\vskip -0.3 cm 
\caption{\small 
Components $\rho_0$ and $\rho_2$, as defined in Eq.~(\ref{eq:rhol}), of the proton (p) and neutron (n) density distributions for the $^{62}$Ni and $^{72}$Se nuclei. The full lines indicate the {\sl optimal solutions} found with the D1S force. Dashed and dotted lines are the distributions of the prolate and oblate solutions, respectively, obtained for the D1ST2a interaction. The {\sl optimal solutions} of the latter two are labelled with a star. The components $\rho_0^{\rm p}$ of the protons are shown in the panels (a) and (e) and the 
$\rho_0^{\rm n}$ of the neutrons in the panels (c) and (g). In the panels (b) and (f) we show the proton $\rho^{\rm p}_2$ components and in panels (d) and (h) the $\rho^{\rm n}_2$ of the neutrons. In the case of the $^{62}$Ni, some of the $\rho_2$ terms  
have been multiplied by the numbers indicated to make them visible at the 
scale of the figure.
}
\label{fig:dens2} 
\end{center} 
\end{figure} 

The effects of the deformation on the density distributions are related to the presence of terms  with $L > 0$ in the expansion of Eq.~(\ref{eq:sumrhol}). In Fig.~\ref{fig:dens2} we show the $L=0$ and $L=2$ components of the proton and neutron HFBCS densities of $^{62}$Ni and $^{72}$Se. In each panel of the figure, we compare the results corresponding to the {\sl optimal solution} obtained with the D1S force (solid curves) with the analogous components of both the prolate (dashed curves) and oblate (dotted curves) solutions obtained with the D1ST2a interaction. We indicate with a star the D1ST2a {\sl optimal solutions}.

At the nuclear surface, all the $\rho_0$ distributions have, essentially, 
the same values. The differences show up in the interior, where the results obtained with the D1S interaction present larger oscillations than the 
other ones. In the case of the $^{72}$Se nucleus 
the behaviour of the $\rho_0$ distributions remains the same, because the 
tensor force only produces a damping of the oscillations (see the dashed curves in Figs. \ref{fig:dens2}e and \ref{fig:dens2}g)). The case of the $^{62}$Ni is more complex. The prolate D1ST2a solution shows proton and neutron densities similar to the D1S ones, even though the oscillations are damped. On the contrary, the densities of the oblate minimum, which is, by the way, the {\sl optimal solution}, present a completely different trend (dotted curves in Figs. \ref{fig:dens2}a and \ref{fig:dens2}c).

The {\sl optimal solution} found for the $^{62}$Ni in the HFBCS calculation with the D1S force is spherical: the corresponding proton and neutron $\rho_2$ components must be multiplied by 100 and 20 to be seen at the scale of the figure (see Figs. \ref{fig:dens2}b and \ref{fig:dens2}d). When the tensor is added, the {\sl optimal solution} becomes oblate, with $\beta_2 \sim -0.1$. 
It is worth pointing out that the solution obtained with the D1ST2a interaction by starting with a set of prolate Wood-Saxon s.p. wave functions (dashed curves in the left panels) has spherical symmetry: it is necessary to use multiplicative factors of 5000 and 1000 in order to show the proton and neutron $L=2$ terms at the scale of the figure.

For the $^{72}$Se nucleus, the {\sl optimal solution} is oblate, with $\beta_2 \sim -0.075$, when the D1S interaction is considered, and changes its shape after including the tensor terms  in the force, becoming a prolate nucleus, with $\beta_2 \sim 0.032$. The $\rho_2$ obtained with the D1ST2a interaction for the oblate solution are of the same order of magnitude than those found with the D1S force, in particular the proton one that remains almost unaltered.

\begin{figure}[!b] 
\begin{center} 
\includegraphics [width=11cm,angle=0]{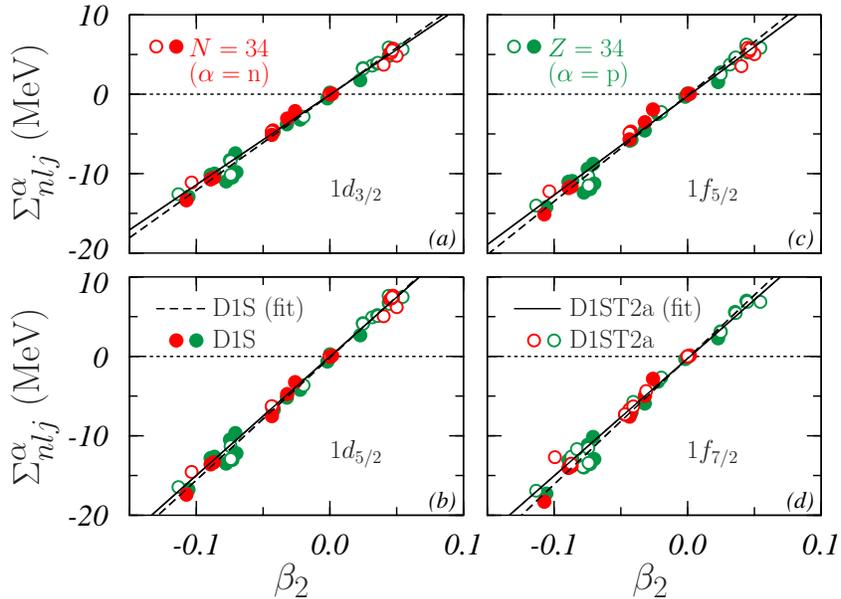} 
\vskip -0.3 cm 
\caption{\small 
Spreads, defined in Eq.~(\ref{eq:spread}), of the s.p. energies of the $1d_{3/2}$, $1d_{5/2}$, $1f_{5/2}$ and $1f_{7/2}$  states vs. the deformation parameter $\beta_2$. The values are those obtained for the {\sl optimal solutions}. The full and empty circles indicate the results found with the D1S and D1ST2a interactions, respectively. The red symbols show the results for the $N=34$ isotone chain and refer to neutron s.p. states. The green symbols, instead, refer to the  $Z=34$ isotope chain and present the results of proton s.p. states. The lines show linear fits to the data: the dashed lines fit the D1S results and the full lines those of the D1ST2a interaction. The values of the fitting parameters are given in Table \ref{tab:fits}.
}
\label{fig:corspread} 
\end{center} 
\end{figure}

\subsection{Single particle energies}
\label{sec:spe}

As said above, in our model, the deformation is obtained by breaking the degeneracy of the s.p. states with the same $n$, $l$, and $j$, quantum numbers since states with different $z$-axis projection $m$ have different energy. Because we assumed time-reversal invariance, those s.p. states with the same $|m|$ remain  degenerated. In order to study the combined effects of the deformation and of the tensor force on the s.p. energies we have considered two quantities: the {\it spread}, $\Sigma^\alpha_{nlj}$, and the {\it centroid splitting}, $\Delta \Gamma^\alpha_{nl}$.

The spread is defined as the difference:
\beq
\Sigma^\alpha_{nlj} \, = \, \epsilon^\alpha_{nlj,m=j}\, - \, \epsilon^\alpha_{nlj,m=1/2} \, , \,\,\,\, \alpha\equiv {\rm p,n} \, .
\label{eq:spread}
\eeq
According to this definition, the  spread is zero for spherical nuclei, positive for prolate solutions and negative for the oblate ones.

We have found a strong relation between the spread of the s.p. energies and the deformation. This is evident in Fig.~\ref{fig:corspread} where we present the s.p. energy spreads of the $1d_{3/2}$, $1d_{5/2}$, $1f_{5/2}$ 
and $1f_{7/2}$ states against the values of $\beta_2$. These results are those of the {\sl optimal solution} for each nucleus considered and have been obtained with the D1S (full circles) and D1ST2a (empty circles) interactions. The values for the nuclei of the $N=34$ isotone chain (red symbols) refer to neutron s.p. states, while those of $Z=34$ isotope chain (green symbols) stand for proton s.p. states. Finally, the straight lines fit the D1S (dashed) and D1ST2a (full) results separately.

It is evident that $\Sigma_{nlj}$ and $\beta_2$ are linearly correlated. 
This is confirmed by the large values of correlation coefficients obtained in the linear fits to the data and shown in Table \ref{tab:fits}. Even more, this correlation is the same independently of the isotonic or isotopic chain analyzed. The slopes of the dashed and full lines are very similar, indicating a small effect of the tensor force. On the other hand, the lines fitting the data of the s.p. states with $j=l+1/2$ (Figs.~\ref{fig:corspread}a and \ref{fig:corspread}c) are steeper than those corresponding to $j=l-1/2$ (Figs. \ref{fig:corspread}b and \ref{fig:corspread}d) as indicated by the values of the coefficients shown in Table \ref{tab:fits}.

\begin{table} [ht]
\begin{center}
\begin{tabular} {cccc c ccc} 
\hline\hline
& \multicolumn{3}{c}{D1S} & ~~~& \multicolumn{3}{c}{D1ST2a} \\ \cline{2-4} \cline{6-8}
quantity & $a$ & $b$ & $r$ & & $a$ & $b$ & $r$ \\ \hline
$\Sigma_{1d3/2}$ & $-0.11\pm 0.20$ & $119.73 \pm 3.30$ & 0.991 & & 
                                 $-0.05\pm  0.11$ & $113.44\pm 2.09$ & 0.996 \\ 
$\Sigma_{1d5/2}$ & $-0.17\pm  0.21$ & $154.18\pm 3.46$ & 0.994 & &
                                 $-0.03\pm 0.13$ & $149.61\pm  2.60$ & 0.996 \\
$\Sigma_{1f5/2}$ & $-0.11\pm 0.25$ & $134.00\pm 4.17$ & 0.989 & &
                                $-0.24\pm  0.14$ & $124.68\pm  2.81$ & 0.994 \\ 
$\Sigma_{1f7/2}$ & $-0.22\pm 0.24$ & $159.03\pm 3.96$ & 0.993 & &
                                $-0.27\pm 0.14$ & $147.84\pm  2.81$ & 0.996\\
\hline
$\Delta \Gamma_{1p}$ & $2.68\pm 0.17$ & $5.03\pm 2.86$ & 0.337 & &
                                       $1.65\pm 0.20$ & $-3.51\pm 3.63$ & -0.194 \\
$\Delta \Gamma_{1d}$ & $4.99\pm 0.21$ & $5.13\pm 3.51$ & 0.286 & &
                                       $3.35\pm 0.27$ & $-5.32\pm 4.96$ & -0.214 \\
$\Delta \Gamma_{1f}$ & $7.71\pm 0.19$ & $1.86\pm 3.20$ & 0.118 & &
                                      $5.80\pm 0.29$ & $-3.97\pm 5.26 $ & -0.152 \\
\hline\hline
\end{tabular}
\caption{Parameters of the linear fits of $\Sigma_{nlj}$ and $\Delta \Gamma_{nlj}$, as a function of $\beta_2$, shown in Figs. (\ref{fig:corspread}) and (\ref{fig:corrcent}). In both cases, the fitting function is $y=a+b\beta_2$. The uncertainties of the parameters and the linear correlation coefficients, $r$, are also given.}
\label{tab:fits}
\end{center} 
\end{table}

The second quantity that we have used in our study of the s.p. energies is the centroid splitting:
\beq
\Delta \Gamma^\alpha_{nl} \,=\, \Gamma^\alpha_{nl,j=l-1/2} \, - \, \Gamma^\alpha_{nl,j=l+1/2} \, ,
\label{eq:so-splitting}
\eeq
where we have indicated with
\beq
\Gamma^\alpha_{nlj} \,=\, \frac{2}{2j\,+\,1} \sum_{m=\half}^{j} \epsilon^\alpha_{nljm} \, , \,\,\,\, \alpha\equiv {\rm p,n} \, ,
\label{eq:centroid}
\eeq
the centroid of the s.p. energies of the multiplet with quantum numbers $n$, $l$ and $j$. 

\begin{figure}[!th] 
\begin{center} 
\includegraphics [width=11cm,angle=0]{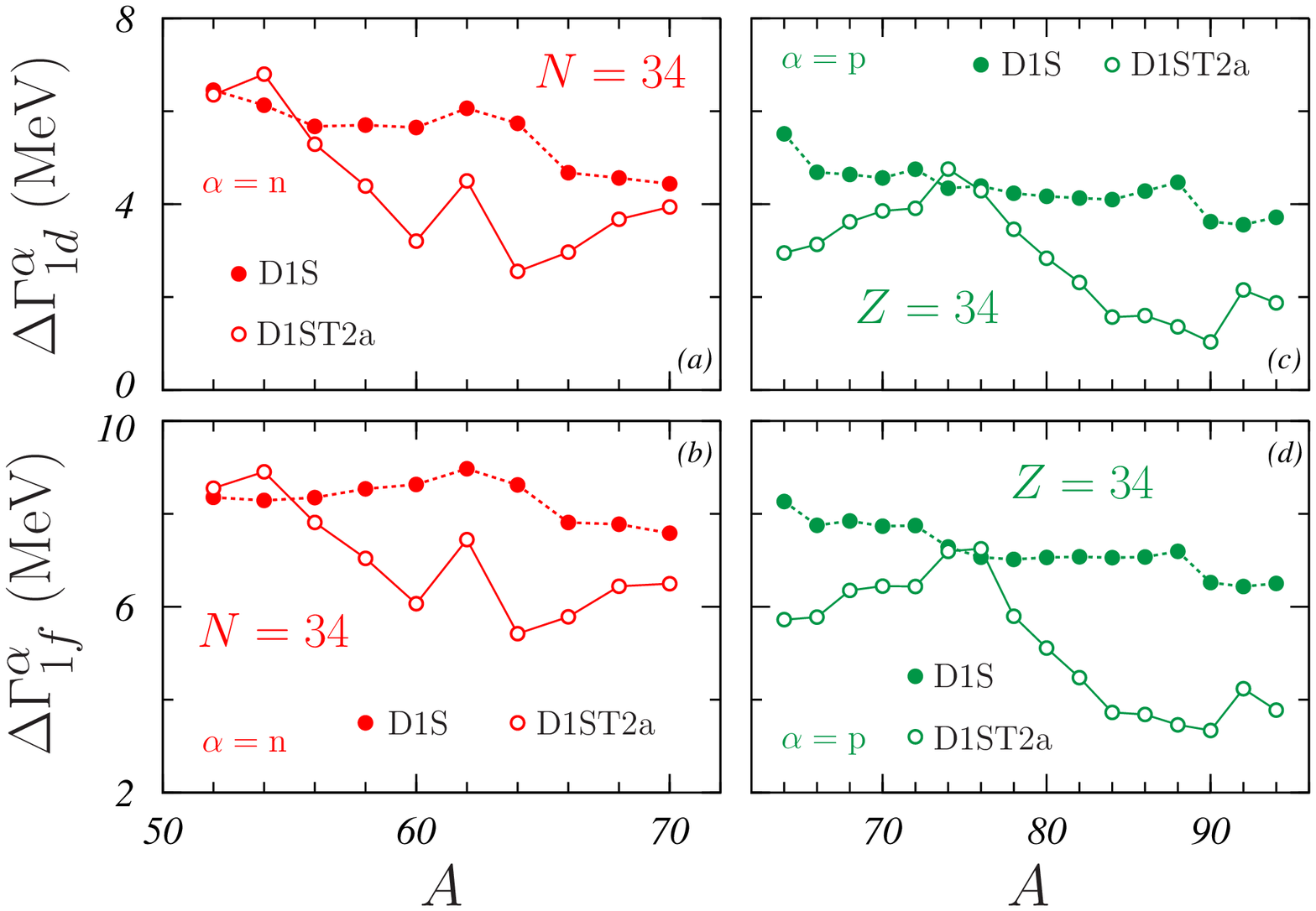} 
\vskip -0.3 cm 
\caption{\small 
Splittings between centroid energies of spin-orbit partner multiplets, Eq.~(\ref{eq:so-splitting}),
obtained in the {\sl optimal solutions}. In the upper panels we show the results  
for the $1d$ states, and in the lower panels those for the $1f$ states. The results of panels (a) 
and (b) are those of the neutron s.p. states of  the $N=34$ isotone chain.
Those of the panels (c) and (d) refer to the proton s.p. states of the $Z=34$ isotope chain.
The full circles indicate the results obtained with the D1S interaction and the empty circles
those of the calculations done with the D1ST2a interaction.
}
\label{fig:centroid} 
\end{center} 
\begin{center} 
\includegraphics [width=6.5cm,angle=0]{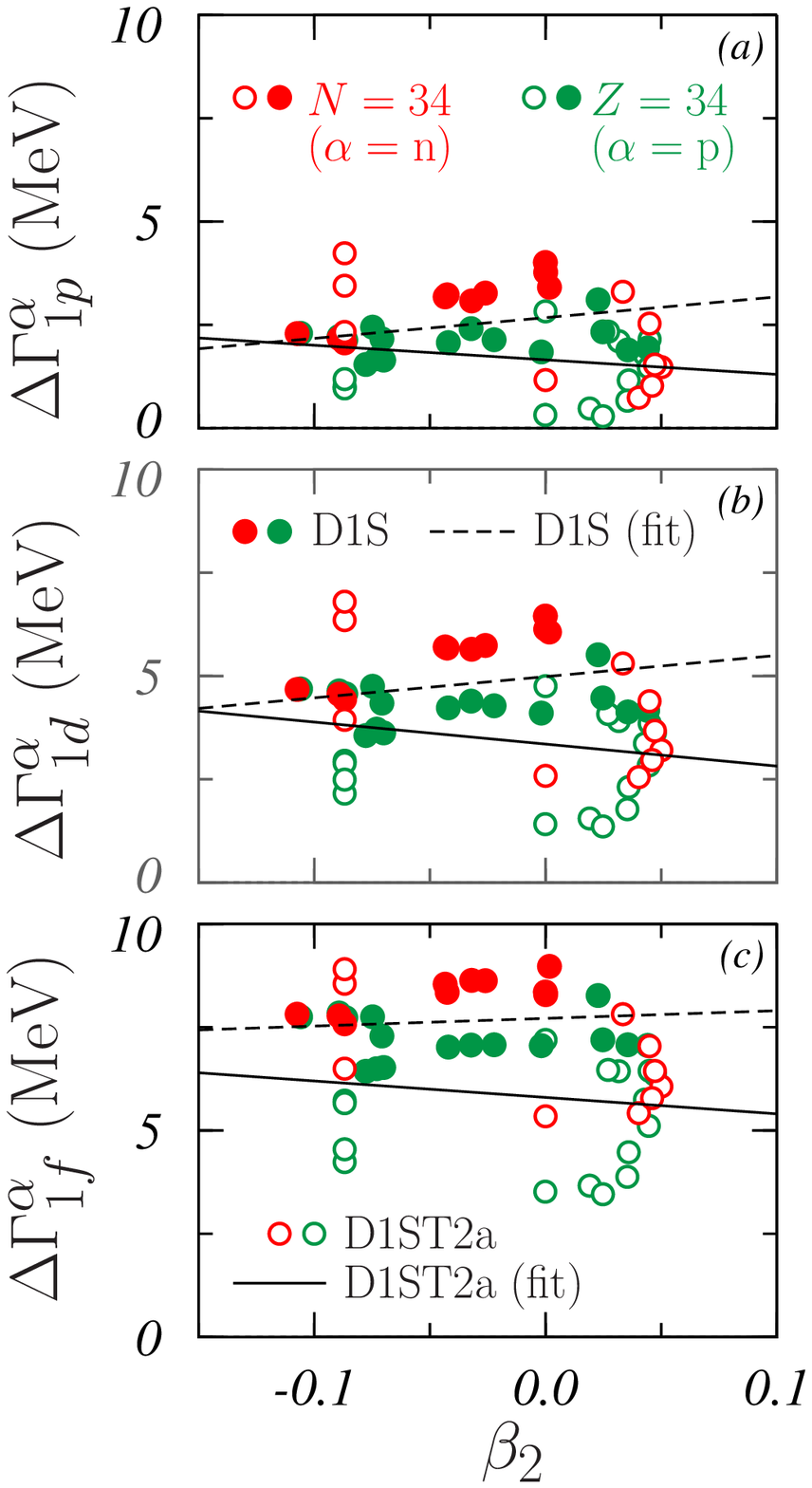}  
\vskip -0.3 cm 
\caption{\small 
Splittings between centroid energies, Eq.~(\ref{eq:so-splitting}), of the 
$1p$ (panel~(a)), 
$1d$ (panel~(b)), and $1f$ (panel~(c)) s.p. states vs. the deformation parameter $\beta_2$.
The results are those of the {\sl optimal solutions}. The red symbols show the results for the
neutrons states of the $N=34$ isotones, the green symbols those of the proton states of 
the $Z=34$ isotopes. The results
obtained with the D1S interaction are indicated by the full circles and those with the D1ST2a
by empty circles. Linear fits to the D1S and D1ST2a results are indicated by the dashed and full lines, respectively. The values of the fitting parameters are given in Table \ref{tab:fits}.
}
\label{fig:corrcent} 
\end{center} 
\end{figure}

In Fig.~\ref{fig:centroid} we show the values of $\Delta \Gamma^\alpha$ obtained in our HFBCS calculations for the $1d$ (upper panels) and $1f$ (lower panels) multiplets. The left panels indicate the results for the neutron s.p. states of the $N=34$ isotones and the right panels those of the proton s.p. states of the $Z=34$ isotopes. The results obtained with the D1S and D1ST2a forces are indicated by the full and empty circles, respectively. 

The tensor force reduces the value of $\Delta \Gamma^\alpha$. This behaviour is similar to the well-known effect that has been pointed out,  discussed and explained for spherical systems by Otsuka and collaborators \cite{ots05,ots06}. In that case, the tensor produces a reduction of the splitting between spin-orbit partners, which is precisely a quantity equivalent to the centroid splitting defined in Eq.~(\ref{eq:so-splitting}) for the deformed nuclei. We have checked that this effect also occurs in the results of the spherical HF+BCS calculations that we have performed for all the nuclei here considered.

\newpage\clearpage

\newpage\clearpage

Since we have found a good correlation between s.p. energy spread and deformation, see Fig.~\ref{fig:corspread}, we repeated an analogous study also for the centroid splitting. We present in Fig.~\ref{fig:corrcent} the values of $\Delta \Gamma^\alpha$ against the deformation parameter $\beta_2$. The results are those obtained for the {\sl optimal solutions} of each nucleus considered, in the cases of the $1p$ (Fig.~\ref{fig:corspread}a), $1d$ (Fig.~\ref{fig:corspread}b), and $1f$ (Fig.~\ref{fig:corspread}c) s.p. states. We show the results for the neutron states in the case of the $N=34$ isotones (red symbols) and for the proton states in the case of the $Z=34$ isotopes (green symbols). Full and empty circles indicate the values obtained by using D1S and D1ST2a interactions, respectively. The data do not show any evident correlation with $\beta_2$. The linear fits of the D1S (dashed lines) and D1ST2a (full lines) data remain almost constant against the changes of the deformation parameter. The absence of correlation between $\Delta \Gamma_{nlj}$ and $\beta_2$ is quantitatively defined by the low values of the correlation coefficients given in Table \ref{tab:fits}.

\begin{figure}[!b] 
\begin{center} 
\includegraphics [width=12cm,angle=0]{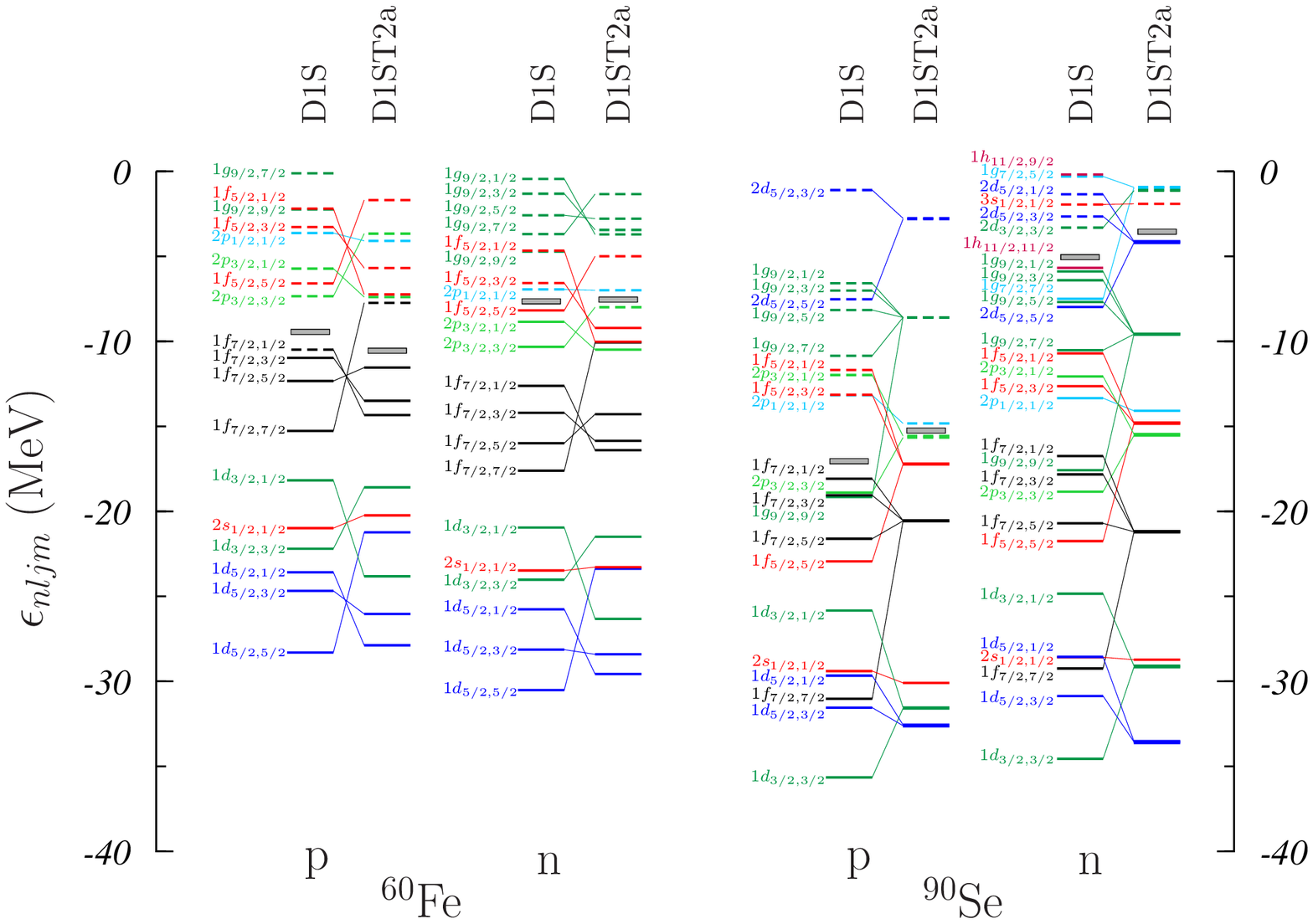}  
\vskip -0.3 cm 
\caption{\small 
Proton (p) and neutron (n) s.p. spectra of the $^{60}$Fe and $^{90}$Se nuclei for the {\sl optimal solutions}
obtained with the D1S and D1ST2a interactions.  The subindexes identify $j$ and $|m|$.
The states with an occupation probability $v^2 <  0.5$ are indicated by dashes lines. 
The thick grey lines show the Fermi levels.
}
\label{fig:spec} 
\end{center} 
\end{figure} 

The tensor terms of the force modify the sequence of s.p. states in the deformed nuclei. As an example of this effect, we show in Fig.~\ref{fig:spec} the proton and neutron s.p. spectra of the {\sl optimal solutions} of 
the $^{60}$Fe and $^{90}$Se nuclei obtained with the D1S and D1ST2a interactions.

The $^{60}$Fe nucleus is oblate for D1S, with $\beta_2=-0.032$, and prolate for D1ST2a, with $\beta_2=0.050$. In the case of the $^{90}$Se nucleus, the {\sl optimal solution} obtained with the D1S force is oblate, with $\beta=-0.07$, while for the D1ST2a interaction we obtain a value of  $\beta_2=8.7\cdot 10^{-5}$, indicating an essentially spherical shape.

The effect of the tensor is quite evident in $^{60}$Fe. There is a slight increase of the spreading for the states with the same $(n,l,j)$ values, and the inversion of the order of the levels with different $|m|$, due to the change of shape. The result of these two combined effects is that for the $1d$ states a level with $j=3/2$, the $1d_{3/2,1/2}$ state, has an energy lower than
a state with $j=5/2$, the  $1d_{5/2,5/2}$ level. 

A said above, the {\sl optimal solution} obtained with the D1ST2a interaction for $^{90}$Se has a spherical shape. This is evident because all the s.p. levels with the same $n$, $l$, and $j$ quantum numbers and different 
$|m|$, converge to an unique energy value. The deformed results obtained with the D1S force show, in some case, an inversion of the levels with different $j$ by placing states with $j=l-1/2$ below states with $j=l+1/2$. This happens for the $1f_{5/2,5/2}$ level whose energy is smaller than that of the three $1f_{7/2,|m|}$ states, and for the the $1f_{7/2,7/2}$ level whose energy lies between those of the $1d_{5/2,1/2}$ and  $1d_{5/2,3/2}$ levels.

Single particle states with their specific characteristics are the basic entities of our model. In order to have the possibility of comparing our predictions with some empirical observation we exploit the Koopman's theorem \cite{rin80} which establishes that, in mean-field models, the global 
properties of odd-even nuclei are fully determined by those of the s.p. states of the unpaired nucleon. 

By considering this approach, we have evaluated the angular momenta and parities of the ground states of those nuclei having one proton less than those considered in the $N=34$ chain of isotones. For the case of the $Z=34$ isotopes, we considered those odd-even nuclei with one neutron less than the even-even isotope partner. The angular momenta and parities of these odd-even nuclei are presented in Table \ref{tab:j-par} and compared to the empirical data taken from the compilation of Ref.~\cite{bnlw}. The parentheses in some of these experimental assignments indicate that they are not fully identified. Since our model considers partial occupations of the s.p. levels, sometimes the definition of the Fermi level is quite ambiguous, and this generates uncertainty in the definition of the last occupied level, which, on the contrary, would be well identified in pure HF calculations. 

By using the data of Table~\ref{tab:j-par} we can better analyze the results of Fig.~\ref{fig:spec}. 
The spin-parity of the ground state of the $^{59}$Mn obtained in our calculation is $7/2^-$ while the experimental value indicated in \cite{bnlw} is $5/2^-$, even though there are uncertainties on the spin assignement. In our calculations, both with and without tensor force, the energies of the proton $1f_{5/2}$ states are always larger than those of the $1f_{7/2}$ states, therefore our model does not account for that spin. From the neutron point of view, we remark that the empirical value of the $^{59}$Fe 
nucleus given in \cite{bnlw} is $3/2^-$, properly predicted by the calculation with the D1ST2a interaction.

The experimental spin-parity assignment of the $^{89}$Se ground state is $(5/2)^-$, correctly described by our calculation performed with the tensor force. Since the spin-parity of the $^{89}$As is unkown, we cannot make a comparison with our predictions for the neutrons.

The results shown in Table~\ref{tab:j-par} indicate that experimental spin-parity assignments are reproduced in only about half of the cases. Our calculations are unable to provide adequate results in all the Se isotopes up to $^{81}$Se with the only exception of the $^{69}$Se nucleus, for the D1S interaction. In nine of the nuclei investigated the results obtained with the D1S and D1ST2a interactions are different.

\begin{table}
\begin{center}
\begin{tabular} {cccc c cccc} 
\hline\hline
\multicolumn{4}{c}{$N=34$} &~~~&  \multicolumn{4}{c}{$Z=34$} \\ \cline{1-4} \cline{6-9}
nucleus & D1S & D1ST2a & exp. && nucleus & D1S & D1ST2a & exp. \\ 
\hline
\rule{0cm}{0.3cm}
$^{51}$Cl & $\ds 3/2^+$ & $\ds 1/2^+,3/2^+$ & &&
$^{65}$Se & $\ds 7/2^-$ & $\ds 7/2^-$ & $\left(\ds 3/2^-\right)$ \\
$^{53}$K & $\ds 3/2^+$ & $\ds 1/2^+$ & $\left(\ds 3/2^+\right)$ &&
$^{67}$Se & $\ds 7/2^-$ & $\ds 5/2^-$ & \\
$^{55}$Sc & $\ds 7/2^-$ & $\ds 7/2^-$ & $\left(\ds 7/2\right)^-$ &&
$^{69}$Se & $\ds 1/2^-$ & $\ds 3/2^-$ & $\ds 1/2^-$ \\
$^{57}$V & $\ds 7/2^-$ & $\ds 7/2^-$ & $\left(\ds 7/2^-\right)$ &&
$^{71}$Se & $\ds 3/2^-$ & $\ds 3/2^-$ & $\left(\ds 5/2^-\right)$ \\
$^{59}$Mn & $\ds 7/2^-$ & $\ds 7/2^-$ & $\left(\ds 5/2\right)^-$ &&
$^{73}$Se & $\ds 5/2^-$ & $\ds 1/2^-$ & $\ds 3/2^-$ \\
$^{61}$Co & $\ds 7/2^-$ & $\ds 3/2^-,7/2^-$ & $\ds 7/2^-$ &&
$^{75}$Se & $\ds 5/2^-$ & $\ds 9/2^+$ & $\ds 5/2^+$ \\
$^{63}$Cu & $\ds 3/2^-$ & $\ds 3/2^-,5/2^-,7/2^-$ & $\ds 3/2^-$ &&
$^{77}$Se & $\ds 9/2^+$ & $\ds 9/2^+$ & $\ds 1/2^-$ \\
$^{65}$Ga & $\ds 9/2^+$ & $\ds 5/2^-$ & $\ds 3/2^-$ &&
$^{79}$Se & $\ds 9/2^+$ & $\ds 9/2^+$ & $\ds 7/2^+$ \\
$^{67}$As & $\ds 7/2^-$ & $\ds 5/2^-$ & $\left(\ds 5/2^-\right)$ &&
$^{81}$Se & $\ds 9/2^+$ & $\ds 9/2^+$ & $\ds 1/2^-$ \\
$^{69}$ Br & $\ds 1/2^-$ & $\ds 1/2^-$ & &&
$^{83}$Se & $\ds 9/2^+$ & $\ds 9/2^+$ & $\ds 9/2^+$ \\
&&&&&
$^{85}$Se & $\ds 5/2^+$ & $\ds 5/2^+$ & $\left(\ds 5/2\right)^+$ \\
&&&&&
$^{87}$Se & $\ds 5/2^+$ & $\ds 5/2^+$ & $\left(\ds 5/2^+\right)$ \\[2mm]
&&&&&
$^{89}$Se & $\ds \frac{11}{2}^-$ & $\ds 5/2^+$ & $\left(\ds 5/2^+\right)$ \\
&&&&&
$^{91}$Se & $\ds 3/2^+$ & $\ds 9/2^+$ & \\
&&&&&
$^{93}$Se & $\ds 5/2^+$ & $\ds 1/2^+,3/2^+,5/2^+$ & $\left(\ds 1/2^+\right)$ \\
\hline\hline
\end{tabular}
\caption{Ground state angular momenta and parities of the odd-even nuclei of the $N=34$ and $Z=34$ chains, obtained according the Koopman's theorem. The experimental values taken from Ref.~\cite{bnlw} are shown for comparison. The values between parentheses indicate that the corresponding assignments are not yet definite.}
\label{tab:j-par}
\end{center}
\vspace*{-0.55cm}
\end{table}


\section{Summary and conclusions}
\label{sec:conclusions}

In this article, we have presented a model describing open shell nuclei. This model is based on the variational principle and uses Slater determinants built with s.p. wave functions whose radial part depends on $m$, the 
projection of the total angular momentum $j$ on the quantisation axis $z$. This feature automatically introduces a deformation in the many-body state. Each step of the iterative procedure minimising the energy functional with these trial wave functions consists of two different calculations. In the first one, we solve the HF equations (\ref{eq:hfradial}) generating the s.p. wave functions, and in the second calculation, we solve a set of BCS equations which modify the occupation probabilities of the s.p. states. Since the solution of HF and BCS equations is considered in each step of  the minimisation procedure, we named HFBCS our model, to distinguish it from the approach of Ref. \cite{ang14} where the solution of the BCS equations is carried out after the full solution of the HF equations has been found. We called HF+BCS this latter approach which, by the way, uses spherical s.p. wave functions. We consistently use the same finite-range interaction to carry out both HF and BCS calculations. We have considered the Gogny type D1S interaction \cite{dec80} and an extension of it, the D1ST2a force, which also contains tensor terms  \cite{ang12}.

The iterative procedure starts from trial wave functions which already have a deformation. In all our calculations we have found that the type of deformation is conserved until convergence is reached. This means that prolate or oblate trial wave functions lead to final results with the same type of deformation. We found the same feature in the HFB calculations of Ref.~\cite{ang01a}.

In general, the two, oblate and prolate, solutions, obtained for each nucleus considered, have very similar total nuclear energies. We have called {\sl optimal solution} that with the smallest energy value. 

The aim of our study was the investigation of how the deformation of the nuclear ground state emerged in our model, the effects on observable  quantities, the effects of the tensor terms  of the effective interaction and the, eventual, relation between deformation and tensor force. We presented results of our HFBCS model regarding energies, density distributions and single particle properties of medium-heavy nuclei 
belonging to the $N=34$ isotone chain and to the Se, $Z=34$, isotope chain. 

Deformation effects on energies and radii are rather small. The total energies obtained in our HFBCS calculations are lower, at most of 1.5\%, with respect to those of the spherical HF+BCS results. The differences between the r.m.s. radii obtained with these two different approaches are even 
smaller.

The effects of the tensor force are more evident. Calculations carried out with the D1ST2a interaction produce nuclei which are slightly less bound than those described by the D1S force. The relative differences between the results of the two calculations are smaller than 2\% for all the cases considered, but the effect is clear and consistent in all nuclei investigated. This effect is worsening the agreement with the experimental energies which are smaller than those obtained without tensor by 1\% at most. All these facts are in compliance with the variational principle which provides upper limits of the correct energy eigenvalues, and with the fact that the global fit to the experimental energies and radii carried out to select the parameters of the interaction has been done for the interaction without tensor. 

Also the effects of the tensor force on the density radii are quite small; relative differences with the results obtained without tensor are smaller than 1\%. These effects have the same sign in almost all the nuclei we have studied. Calculations carried out with tensor terms in the interaction produce r.m.s. radii larger than those obtained without them. 

A detailed investigation of proton, neutron and charge density distributions has shown that the most evident differences between the results of the various calculations show up in the interior of the nucleus. Densities without tensor present a rather oscillating behaviour in the nuclear interior. The tensor is smoothing these oscillations and, for the cases where results are available, we found a better agreement with the empirical charge densities. 

The s.p. energies are the quantities most affected by deformation and tensor force which, both, generate a reordering of the s.p. level scheme. The deformation destroys the $2j+1$ degeneracy of the spherical s.p. states 
characterised by the $n$, $l$ and, obviously, $j$ quantum numbers. We have assumed rotational symmetry around the $z$ axis, and time-reversal symmetry, therefore, we obtain different s.p. energies for each value of $|m|$. Each spherical, and $2j+1$ degenerated, s.p. state is split in $j+1/2$ 
different states. We have defined the {\it spread} as the difference between the s.p. energies of the two extreme states with the same $j$ (those with $|m|=j$ and $|m|=1/2$) and we found a strong linear correlation between its values and those of the deformation parameter $\beta_2$. This 
correlation is present for both oblate and prolate solutions obtained with or without tensor force. 

The tensor force changes the type of deformation of the {\sl optimal solution}, therefore the last occupied proton or neutron s.p. state. A comparison between the measured angular momenta of odd-even nuclei and those of 
the last occupied states does not show any specific trend and does not provide a real preference between calculations carried out with or without tensor. The only clear facts are that the two type of calculations produce different results in some of the cases analyzed, and only in half of the nuclei considered one the two calculations is able to predict the experimental values. 

The parameter which better summarizes the information on the deformation is $\beta_2$, defined in Eq.~(\ref{eq:beta2}). Our calculations generate $\beta_2$ values remarkably smaller, in absolute value, than those obtained in HFB calculations. Also the comparison with the values obtained by an empirical model \cite{mol16} and those indicated as experimental data \cite{bnlw} shows that our results are smaller, in absolute value. The size of the deformation is essentially the same for calculations carried out 
with and without tensor force, even though, in general, the {\sl optimal solutions} with tensor force are less deformed than those without it. 

The results of our study clearly indicate that the present accuracy of the experimental data on binding energies, charge radii and distributions imposes a new global fit of a force containing tensor terms in such a way that all the force parameters will be modified.

Our HFBCS approach proposes a peculiar manner to describe open-shell nuclei which automatically generates deformations in nuclear ground states. Under many aspects this approach is simpler than that of the HFB model, and shows s.p. properties that are still well recognizable. The extension of this approach to describe odd-even nuclei is under way. The set of s.p. wave functions with their occupation probabilities is the starting point to build up a Deformed Quasi-Particle Random Phase Approximation.

\acknowledgments  
This work has been partially supported by the Junta de Andaluc\'{\i}a (FQM387), the Spanish Ministerio de 
Econom\'{\i}a y Competitividad (PID2019-104888GB-I00) and the European 
Regional Development Fund (ERDF). 

\appendix
\section{The HF potential terms }
\label{sec:hfmatel}

We express the effective nucleon-nucleon interaction in terms  of operator channels as it is done in the Argonne-Urbana potentials \cite{wir95}:
\beqn
V(r_{12}) &=& \sum_{p=1}^6 V_p(r_{12}) \, O^p(1,2) \,+\, V_{\rm Coul}(r_{12}) 
\, \frac{\left[ 1 + \tau_z(1)  \right] \, \left[ 1 + \tau_z(2)  \right]}{4} \, 
\label{eq:int}
\\ \nonumber &&
+ \, V_{LS} \, {\bf L}_{12} \cdot {\bf S}_{12} \, \delta(\br_1 - \br_2)  \,
+ \, [ V^c_\rho \,+\, V^s_\rho \, \bsigma(1)  \cdot \bsigma(2) ] \, P(\rho)  \, \delta(\br_1 - \br_2) 
\, .
\eeqn
Here $r_{12}=|\br_1 \,-\, \br_2|$ is the distance between the two interacting nucleons and
the six operators $\left\{ O^p,\, p=1,\ldots,6 \right\}$ are, respectively, 
\beq
1, \,\btau(1) \cdot \btau(2),\, \bsigma(1) \cdot \bsigma(2),
\, \bsigma(1) \cdot \bsigma(2)\, \btau(1) \cdot \btau(2),\, 
S_{12},\, S_{12}\, \btau(1) \cdot \btau(2) \, ,
\label{eq:operator}
\eeq
with $\bsigma(i)$ the Pauli operator corresponding to the spin of the $i$-th nucleon and $S_{12}$ the tensor operator, which we define as
\beq
S_{12} \,=\, 3\, \frac{ [\bsigma(1) \cdot \br_{12}]\, [\bsigma(2) \cdot 
\br_{12}] }{\br^2_{12}}
\,-\, \bsigma(1) \cdot \bsigma(2) 
\label{eq:tens}
\, .
\eeq
In Eq. (\ref{eq:int}), $\tau_z =1$ for protons and 0 for neutrons, ${\bf L}_{12}$ is the total angular momentum of the nucleonic interacting pair, and ${\bf S}_{12}$ its total spin. Finally, $P(\rho)$ is a scalar function of the nuclear density which we define below, in Eq. (\ref{eq:Prho}).

Using the expression of the operators and potential terms in the momentum 
space, $\widetilde{O}^p(1,2)$ and $\widetilde{V}_p({\bf q})$, respectively, we handle the finite-range part of the interaction by considering the inverse Fourier transform
\beq
V_p(r_{12}) \, O^p(1,2) \,= \, \frac {1}{(2 \pi)^{3/2}} \, \int {\rm d}^3 q \, \exp\left[i\,(\br_1 - \br_2) \cdot {\bf q}\right] \, \widetilde{V}_p({\bf q}) \, \widetilde{O}^p(1,2) \, ,
\eeq
separating the two exponentials and expanding them in multipoles. For the 
four central channels, only the zeroth-order spherical terms  contribute, 
while for the two tensor channels ($p=5,6$), the only term of the expansion contributing is that of the second order. Then it is useful to define the integrals:
\beq
{\cal I}_p(r_1,r_2,L_1,L_2) \, =\, \int_0^\infty\, {\rm d}q \, q^2 \, j_{L_1}(qr_1) \, j_{L_2}(qr_2) \, \widetilde{V}_p(q) \, .
\eeq
Here $j_L$ indicates the spherical Bessel function of $L$-th order. 

Taking into account Eqs.~(\ref{eq:spwave})-(\ref{eq:spwave-ang}), and using the short notation $\hat{j} = \sqrt{2j+1}$ for the indexes indicating angular momenta, we obtain the expressions of the various contributions 
to the potentials ${\cal U}$, ${\cal W}$ and ${\cal K}$. The direct, Hartree, potential ${\cal U}$ for the terms  $p=1,2$ can be expressed as:
\beqn
\nonumber 
{\cal U}^{p=1,2}_{k}(r_1) &=& \sqrt{\frac 2 \pi} \,
\sum_{i} \sum_{L}  v^2_{i} \,
(-1)^{m_k+m_i+1} \, \hat{j_k}^2 \,\hat{j_i}^2 \, \hat{L}^2 \, \xi(L) \,
\int {\rm d}r_2 \,r_2^2 \, R^2_{i}(r_2) \, {\cal I}_p(r_1,r_2,L,L)  \\
&~& 
 \threej{j_k}{L}{j_k}{-m_k}{0}{m_k} \threej{j_k}{L}{j_k}{\half}{0}{-\half} 
 \threej{j_i}{L}{j_i}{-m_i}{0}{m_i} 
 \threej{j_i}{L}{j_i}{\half}{0}{-\half}  \, {\cal T}^p_{\cal U}\, ,
\eeqn
with $\xi(n)=1$ if $n$ is even, and 0 otherwise, and 
\beq
 {\cal T}^p_{\cal U}\, = \, \left\{ 
\begin{array}{ll}
1\, , & p=1 \, ,\\
4\, t_k \,  t_i \, , & p=2\, .
\end{array}
\right.
\eeq
Here the $t$'s represent the eigenvalue of the isospin third component operator that we have chosen to be $1/2$ for protons and  $-1/2$ for neutrons. All the other terms  of the interaction do not contribute to ${\cal U}$.

The contributions of the various interaction terms  to the exchange, Fock-Dirac, potential ${\cal W}$ can be expressed as:
\beqn
\nonumber
{\cal W}^{p=1,2}_k(r_1,r_2) &=& \sqrt{\frac{2}{\pi}}\, \sum_{i} \sum_{L} 
v^2_{i} \, \hat{j_k}^2 \, \hat{j_i}^2 \, \hat{L}^2 \, \xi(l_k+l_i+L) \, R^*_{i}(r_1) \, R_{i}(r_2) \, {\cal I}_p (r_1,r_2,L,L)
 \\
&~& 
\threej{j_i}{L}{j_k}{-m_i}{-M}{m_k}^2\, \threej{j_i}{L}{j_k}{\half}{0}{-\half}^2 
 \, {\cal T}^p_{\cal W}\, , \\[2mm]
%
 {\cal W}^{p=3,4}_k(r_1,r_2) &=& 6 \, \sqrt{\frac{2}{\pi}} \,\sum_{i} 
 \sum_{L\,J} 
v^2_{i} \, \hat{l_k}^2 \,\hat{l_i}^2 \,\hat{j_k}^2 \,\hat{j_i}^2 \, \hat{L}^2 \,\hat{J}^2
\, R^*_{i}(r_1) \, R_{i}(r_2) \, {\cal I}_p (r_1,r_2,L,L) \nonumber \\
&~& 
\threej{l_k}{l_i}{L}{0}{0}{0}^2 \, \threej{j_k}{J}{j_i}{-m_k}{M}{m_i}^2 \,
\ninej {l_k}{\half}{j_k}  {l_i}{\half}{j_i} {L}{1}{J}^2  \, {\cal T}^p_{\cal W} \, , \\
%
\nonumber
{\cal W}^{p=5,6}_k(r_1,r_2) &=& 12 \, \sqrt{\frac{5}{3 \pi}}  \, 
 \sum_{i} \sum_{L_1\, L_2 \, J}
v^2_{i} \,(-1)^J \, (-i)^{L_1-L_2}\, 
\hat{l_k}^2 \, \hat{l_i}^2 \, \hat{j_k}^2 \, \hat{j_i}^2 \,  
\hat{L_1}^2 \,  \hat{L_2}^2  \, \hat{J}^2 \, R^*_{i}(r_1) \, R_{i}(r_2)  \\
 &~& 
{\cal I}_p (r_1,r_2,L_1,L_2) 
 \threej{l_k}{l_i}{L_1}{0}{0}{0} \, \threej{l_k}{l_i}{L_2}{0}{0}{0}
 \threej{j_k}{J}{j_i}{-m_k}{M}{m_i}  \, 
 \nonumber \\
 &~& 
\sixj{L_1}{L_2}{2}{1}{1}{J}
\, \threej{L_1}{L_2}{2}{0}{0}{0} \,
  \ninej {l_k}{\half}{j_k}  {l_i}{\half}{j_i} {L_1}{1}{J} \, 
  \ninej {l_k}{\half}{j_k}  {l_i}{\half}{j_i} {L_2}{1}{J}
 \, {\cal T}^p_{\cal W} \, .
\eeqn
In the previous equations, the isospin term is given by:
\beq
{\cal T}^p_{\cal W} \, = \, \left\{ 
\begin{array}{ll}
 \delta_{t_k,t_i} \, , & p=1,3,5 \, ,\\
2  \,- \, \delta_{t_k,t_i} \, , & p=2,4,6\, .
\end{array}
\right.
\eeq

The contributions of the Coulomb force are, for the Hartree term,
\beqn
\nonumber 
 {\cal U}^{\rm Coul}_{k}(r_1) &=& e^2 \, 
\sum_{i} \sum_{L}  v^2_{i} \, (-1)^{m_k+m_i+1} \, \hat{j_k}^2  \, \hat{j_i}^2 \, \xi(L) 
\, \int_0^\infty {\rm d}r_2 \, r_2^2 \, \frac{r_<^L}{r_>^{L+1}}\, R_i^2(r_2) \\
&~& \nonumber
\threej{j_k}{L}{j_k}{-m_k}{0}{m_k} 
\threej{j_k}{L}{j_k}{\half}{0}{-\half} 
\threej{j_i}{L}{j_i}{-m_i}{0}{m_i} 
\threej{j_i}{L}{j_i}{\half}{0}{-\half} \,\delta_{t_k,t_i}\, \delta_{t_k,\half} \, ,
\eeqn
and, for the Fock-Dirac term,
\beqn
\nonumber
{\cal W}^{\rm Coul}_{k} (r_1,r_2) 
\nonumber &=& e^2 \,\sum_{i}  \sum_{L}  v^2_{i} \,
  \hat{j_k}^2 \, \hat{j_i}^2 \, \xi(l_k+l_i+L) \,
\frac{r_<^L}{r_>^{L+1}} \,  R^*_{i}(r_1) \,R_{i}(r_2) \\
&~& 
 \, \threej{j_i}{L}{j_k}{-m_i}{-M}{m_k}^2 \, \threej{j_i}{L}{j_k}{\half}{0}{-\half}^2
  \,\delta_{t_k,t_i}\, \delta_{t_k,\half} \, .
\eeqn
Here $e$ is the elementary charge and we have indicated, respectively, with $r_<$ and $r_>$ the smaller and the larger values between $r_1$ and $r_2$.

The two zero-range components of our interaction are the density dependent term and the spin-orbit one. The direct, ${\cal Z}\equiv {\cal U}$, and 
exchange, ${\cal Z}\equiv {\cal W}$,  contributions of the latter are given by
\beqn
{\cal Z}_k^{LS}(r_1) &=& 
\frac {1}{4 \pi} \,  V_{LS} \, \sum_i \sum_{\mu_i,s_i} v_i^2 \, \hat{j_i}^2 \, \threej{l_i}{\half}{j_i}{\mu_i}{s_i}{-m_i}^2 
\, {\cal T}_{\cal Z}^{LS}\\ &~&  \left\{ 
\left[j_k(j_k+1) - l_k(l_k+1) - \frac{3}{4}\right] \, \frac{1}{r_1}\, R_{i} (r_1)\, \frac{{\rm d}R_{i}(r_1)}{{\rm d}r_1} \right. \nonumber \\
&~&  \left. - \left[j_i({j_i}+1) - l_i({l_i}+1) - \frac{3}{4} \right] \, 
 \left[ \frac{1}{r}\, R_{i} (r_1)\, \frac{{\rm d}R_{i}(r_1)}{{\rm d}r_1} \,- \,\frac{1}{2\,r_1} \, R^2_{i}(r_1) \right] \right\} \, , \nonumber
\eeqn
where
\beq
{\cal T}^{LS}_{\cal Z} \, = \, \left\{ 
\begin{array}{ll}
1\, , & {\cal Z}\equiv {\cal U} \, , \\
 \delta_{t_k,t_i} \, , & {\cal Z}\equiv {\cal W} \, .
\end{array}
\right.
\eeq

In the case of the density dependent term we have used for the function $P(\rho)$ the expression:
\beq
P(\rho)=\left[ \frac {\rho_0(r_1)\, + \,\rho_0(r_2)} {2} \right]^\gamma 
\, ,
\label{eq:Prho}
\eeq
where $\rho_0(r) = \rho^{\rm p}_0(r) + \rho^{\rm n}_0(r)$ is defined in 
Eq.~(\ref{eq:rhol}).
%
The contributions of this density dependent term can be expressed as:
\beq
{\cal U}^\rho_k(r_1) \, = \, V^c_\rho \, [\rho_0(r_1)]^{\gamma+1}  \, , 
\\
\eeq
%
for the direct potential,
\beq
{\cal W}^\rho_k(r_1,r_2) \,=\, \left( V^c_\rho \,+\, 3\, V^\sigma_\rho \right) \, \left[ \displaystyle \frac{1 \,+\,t_k}{2}\,\rho_0^{\rm p}(r_1) 
\, + \, 
\frac{1 \,-\,t_k}{2}\,\rho_0^{\rm n}(r_1) 
 \right] \, [\rho_0(r_1)]^\gamma \, \delta(r_1-r_2) \,, 
\eeq
for the exchange one, and
\beq
{\cal K}^\rho(r_1) \,=\, \gamma \, [\rho_0(r_1)]^{\gamma-1} \, \left\{ V^c_\rho\,[\rho_0(r_1)]^2\,
-\,\left( V^c_\rho \, + \, 3 \, V^\sigma_\rho \right)\, 
\frac{ [\rho^{\rm p}_0 (r_1)]^2 \, +\, [\rho^{\rm n}_0(r_1)]^2 }{2} 
\right\}  \, ,
\eeq
for the genuine density-dependent term of Eq.~(\ref{eq:K}).

%
%

\section{The BCS matrix elements}
\label{sec:bcsmatel}

In our calculations, the Coulomb, spin-orbit and density dependent terms  
of the interaction (\ref{eq:int}) are not considered in the BCS equations. The pairing interaction that we use is:
\beq
V_{\rm pairing}(r_{12}) \, =\, \sum_{p=1}^6 V_p(r_{12}) \, O^p(1,2) \, .
\label{eq:int-pr}
\eeq
Since we consider the pairing to be active only between like particles, 
the expressions of the interaction terms of Eq.~(\ref{eq:delta})
differing only by the isospin operator $\btau(1) \cdot \btau(2)$ 
are identical. Taking this into account, we have:
\beq
\Delta_{k} = \sum_{p=1}^6 \Delta^p_{k} \, ,
\label{eq:deltak1}
\eeq
where the scalar terms  are
\beqn
\Delta^{p=1,2}_{k} &=& \sqrt{ \frac{2}{\pi} } \, \sum_{i} \sum_L (-1)^L 
\hat{j_k}^2 \, \hat{j_i}^2 \, \hat{L}^2  \, \xi(l_k + l_i +L) 
\\ &~&\nonumber
\int {\rm d}r^2_1 \,{\rm d}r^2_2 \, R_{k}(r_1) \,R_{k}(r_2) \, 
R_{i}(r_1) \, R_{i}(r_2) \, {\cal I}_{p}(r_1,r_2,L,L)
\\ &~& \nonumber
\threej{j_k}{L}{j_i}{m_k}{M}{-m_i}^2 \,\threej{j_k}{L}{j_i}{\half}{0}{-\half}^2 \, ,
\eeqn
the spin terms  can be expressed as
\beqn
 \Delta^{p=3,4}_{k} &=&6 \, \sqrt{ \frac{2}{\pi} } \, 
\sum_{i} \sum_{L\,J} (-1)^{L+1} \,  
\hat{l_k}^2 \, \hat{l_i}^2  \, \hat{j_k}^2 \, \hat{j_i}^2 \,\hat{L}^2 \,\hat{J}^2  
\\ &~& \nonumber 
\int {\rm d}r^2_1 \,{\rm d}r^2_2 \,R_{k}(r_1) \, R_{k}(r_2) \,
R_{i}(r_1)\, R_{i}(r_2) \, {\cal I}_{p}(r_1,r_2,L,L)
\\ &~& \nonumber
\threej{l_k}{L}{l_i}{0}{0}{0}^2  
\threej{j_k}{J}{j_i}{\half}{0}{-\half}^2  
\ninej {l_k} {\half} {j_k} {l_i} {\half} {j_i} {L}{1}{J}^2 \, .
\eeqn
and, finally, the tensor terms  are
\beqn
\nonumber 
\Delta^{p=5,6}_{k} &=&  \frac{3 \sqrt{30} }{\pi^2} \,
\sum_{i} \sum_{L_1\, L_2\, J} 
(-i)^{L_1-L_2} \, (-1)^{L+J+1} \, 
\hat{l_k}^2 \, \hat{l_i}^2 \, \hat{j_k}^2 \, \hat{j_i}^2 \, \hat{L_1}^2 \, \hat{L_2}^2 \, \hat{J}^2 
\\ &~&
\int {\rm d}r^2_1\, {\rm d}r^2_2\, R_{k}(r_1)\, R_{k}(r_2) \, 
R_{i}(r_1) \,R_{i}(r_2)\, {\cal I}_{p}(r_1,r_2,L_1,L_2)
\\ &~& \nonumber
\threej{l_k}{L_1}{l_i}{0}{0}{0} \, \threej{l_k}{L_2}{l_i}{0}{0}{0} \, \threej{L_1}{L_2}{2}{0}{0}{0} \, 
\sixj{L_1}{L_2}{2}{1}{1}{J}
\\ &~& \nonumber
\threej{j_k}{J}{j_i}{m_k}{M}{-m_i}^2 \,  
\ninej{l_k}{\half}{j_k} {l_i}{\half}{j_i}{L_1}{1}{J}  \,
\ninej{l_k}{\half}{j_k} {l_i}{\half}{j_i}{L_2}{1}{J}  \, .
\eeqn

%

\end{document}